\documentclass[
 reprint,
 amsmath,amssymb,
 aps,
prd,
]{revtex4-1}

\usepackage{graphicx}
\usepackage{dcolumn}
\usepackage{bm}
\usepackage{subfigure}
\usepackage{breqn}
\usepackage{lipsum}
\usepackage{etoolbox}
\usepackage{xcolor}

\begin{document}

\newcommand{\trace}{\mathrm{Tr}}
\newcommand{\identity}{{\bf I}}
\newcommand{\SU}[1]{\mathrm{SU}(#1)}
\newlength{\RvQwidth}
\setlength{\RvQwidth}{0.65\columnwidth}

\newcommand{\UM}{{\bf U}}
\newcommand{\VM}{{\bf V}}
\newcommand{\UMC}{{\bf U}^{*}}
\newcommand{\VMC}{{\bf V}^{*}}
\newcommand{\US}{\mathrm{u}}
\newcommand{\VS}{\mathrm{v}}
\newcommand{\USC}{\mathrm{u}^{*}}
\newcommand{\VSC}{\mathrm{v}^{*}}

\makeatletter
\preto\maketitle{%
  \begingroup\lccode`~=`,
  \lowercase{\endgroup
  \let\saved@breqn@active@comma~
  \let~}\active@comma 
}
\appto\maketitle{%
  \begingroup\lccode`~=`,
  \lowercase{\endgroup
  \let~}\saved@breqn@active@comma 
}
\makeatother

\title{Connection between centre vortices and instantons through gauge-field smoothing}

\author{Daniel Trewartha}
\email{daniel.trewartha@adelaide.edu.au}
\author{Waseem Kamleh}
\author{Derek Leinweber}
\affiliation{Centre for the Subatomic Structure of Matter(CSSM), \\
	Department of Physics, University of Adelaide 5005, Australia}

\date{\today}

\begin{abstract}
A recent lattice study of the Landau-gauge overlap quark propagator has shown a connection between centre vortices and dynamical chiral symmetry breaking in $\SU{3}$ gauge theory. We further investigate this relationship through an exploration of the connection to the instanton degrees of freedom. After identifying centre vortices on the lattice in maximal centre gauge, we smooth configurations using multiple algorithms. We are able to create an instanton liquid-like background on configurations consisting solely of centre vortices, analogous to that found on Monte-carlo generated configurations after similar smoothing. Through calculations of the static quark potential and Landau-gauge overlap propagator, we show that this background is able to reproduce all salient long-range features of the original configurations. Thus we conclude that the information necessary to recreate the long-range structure of $\SU{3}$ gauge theory is contained within the centre vortex degrees of freedom.
\end{abstract}


\maketitle


\section{Introduction}

It has long been known that a centre vortex \cite{'tHooft:1977hy,'tHooft:1979uj,Cornwall:1979hz,Nielsen:1979xu,Ambjorn:1980ms,Vinciarelli:1978kp,Yoneya:1978dt,Mack:1978rq} model of the QCD vacuum is capable of explaining confinement, and recent work \cite{Greensite:2014gra} hints that this is the model of confinement most consistent with lattice results. In $\SU{2}$ lattice gauge theories, centre vortex gauge configurations have been found to recover the full string tension, and thus confinement \cite{DelDebbio:1998uu,DelDebbio:1996mh}. Centre vortices have also been found to be responsible for dynamical chiral symmetry breaking \cite{Bowman:2008qd,deForcrand:1999ms,Alexandrou:1999vx,Engelhardt:2002qs,Bornyakov:2007fz,Hollwieser:2008tq,Hollwieser:2013xja,Hollwieser:2014osa,Hoellwieser:2014isa,Alexandrou:1999vx,Kovalenko:2005rz}, and thus are the dominant long-range structure of the $\SU{2}$ vacuum. \par 
In $\SU{3}$ gauge theory, however, fewer studies exist, and so the picture is less clear. While removal of centre vortices still removes the string tension -- and thus confinement -- a background consisting solely of centre vortices reproduces just $66\%$ of the original string tension \cite{Langfeld:2003ev,Cais:2008za}. Additionally, while studies of dynamical chiral symmetry breaking through the Landau-gauge AsqTad quark propagator were unable to show a loss of dynamical chiral symmetry breaking coincident with vortex removal \cite{Bowman:2010zr}, this did occur in a study of the low-lying hadron spectrum \cite{OMalley:2011aa}. Recently, the work of Ref.~\cite{Trewartha:2015nna}, studying the Landau-gauge quark propagator using the chirally sensitive overlap fermion action, revealed removal of dynamical chiral symmetry breaking following vortex removal, and furthermore reproduction of dynamical mass generation after small amounts of cooling on configurations consisting solely of centre vortices. \par
These results raise the possibilty that previous studies of centre vortex configurations in $\SU{3}$ gauge theory suffered from a lack of smoothness; since gauge field configurations consisting solely of centre vortices have links consisting entirely of centre elements of $\SU{3}$ by definition, they are high action, rough configurations. In this work we will examine the effects of smoothing vortex-only configurations, comparing to smoothed Monte Carlo generated (``untouched'') configurations. We find a reproduction of the instanton-like degrees of freedom on vortex-only configurations, providing a mechanism for the dynamical mass generation seen in Ref.~\cite{Trewartha:2015nna}. \par
We begin in section \ref{sec:VortexIdent} with a brief overview of the process of identifying centre vortices on the lattice. In section \ref{sec:smoothing} we examine the instanton content of these configurations through direct examination of the gauge field. We perform the smoothing procedures of cooling and over-improved stout link smearing on the configurations, and examine the resulting instanton-like objects. Instantons are associated with low-lying modes of the Dirac operator \cite{Ilgenfritz:2008ia}, and thus through the Banks-Casher relation \cite{Banks:1979yr} provide a mechanism for dynamical chiral symmetry breaking. We show an essentially equivalent background of instanton-like objects emerging on both vortex-only and untouched configurations through the process of smoothing. In section \ref{sec:SQP} we analyse heavy quark confinement using the static quark potential, and find, again, almost perfect agreement between vortex-only and untouched configurations after smoothing. In section \ref{sec:QProp} we analyse dynamical mass generation, and thus dynamical chiral symmetry breaking, through the Landau-gauge overlap quark propagator, showing that vortex-only configurations are able to reproduce dynamical mass generation at all levels of smoothing considered, over a range of bare quark masses.

\section{Identifying Centre Vortices on the Lattice}
\label{sec:VortexIdent}

The strategy for identifying thick centre vortices on the lattice is simple; we seek to decompose gauge links in the form
\begin{equation}
U_{\mu}(x) = Z_{\mu}(x)\,R_{\mu}(x),
\end{equation}
where $Z_{\mu}(x)$ is an element of the centre of $\SU{3}$,
\begin{equation}
Z_{\mu}(x) = \exp \left[ \frac{2\pi i}{3}m_{\mu}(x) \right] \,\identity,
\end{equation}
in such a way that all vortex information is captured in the centre-projected elements $Z_{\mu}(x)$, and $R_{\mu}(x)$ thus containing the remaining short-range noise. The centre flux through plaquettes $P_{\mu\nu}(x)$ is identified via their centre projection,
\begin{equation}
P_{\mu\nu}(x) = Z_{\mu}(x)\,Z_{\nu}(x+\mu)\,Z_{\mu}^{\dagger}(x+\nu)\,Z_{\nu}^{\dagger}(x),
\end{equation}
where there exists a non-trivial centre flux if $P_{\mu\nu}(x) \neq \identity$. The set of such plaquettes form closed surfaces; the ``thin'' centre vortices. These are embedded within the thick centre vortices of the original, Monte Carlo generated, configurations. \par
The natural choice \cite{DelDebbio:1998uu} to perform this decomposition is to find some gauge transformation $\Omega(x)$ which minimizes the distance between links and centre elements; i.e.,
\begin{equation}
\sum_{x,\mu}\,||U_{\mu}^{\Omega}(x) - Z_{\mu}(x)|| \quad \rightarrow \quad \textrm{min,}
\end{equation}
noting that we minimize over both the gauge transformation $\Omega(x)$ and choice of centre element $Z_{\mu}(x)$. $U_{\mu}^{\Omega}(x)$ denotes the link $U_{\mu}(x)$ after performing the gauge transformation $\Omega(x)$. Choices of gauge fixing functional fulfilling this condition are collectively known as Maximal Centre Gauge. Since all centre elements are mapped to the identity in the adjoint representation, this is equivalent to maximising the trace of the links in the adjoint representation. There are many possible choices \cite{DelDebbio:1998uu,DelDebbio:1996mh,Vink:1992ys,Vink:1994rb,Alexandrou:1999vx,Faber:2001zs,Montero:1999by} of gauge fixing functional which achieve this. We use the so-called ``mesonic'' maximal centre gauge \cite{Montero:1999by},
\begin{equation}
R = \frac{1}{V\,N_{dim}\,3^{2}}\sum_{x,\mu}\,|\trace \, U_{\mu}^{\Omega}(x)|^{2} \quad \rightarrow \quad \textrm{max},
\end{equation}
where the normalisation factor, $1 /\, VN_{dim}\,3^{2}$, is introduced to guarantee $|R| \le 1$. Setting $\Omega(x) = \identity$ everywhere except at a single lattice site $x'$, we locally maximize the quantity
\begin{eqnarray}
R_{\textrm{local}}(x') = &\sum_{\mu}&|\, \trace \, \Omega(x')\,U_{\mu}(x')|^{2} \,+\,  \\ \nonumber
&\sum_{\mu}&|\, \trace \, U_{\mu}(x' - \hat{\mu})\,\Omega^{\dagger}(x')|^{2}.
\end{eqnarray}
We restrict $\Omega(x')$ to be in one of the $\SU{2}$ subgroups of $\SU{3}$. We parametrize an $\SU{2}$ matrix $[\Omega(x')]_{\SU{2}}$ by
\begin{equation}
[\Omega(x')]_{\SU{2}} = \Omega_{4}\,\identity - i\Omega_{i}\,\sigma_{i},
\end{equation}
and embed the resulting $\SU{2}$ matrix in one of the 3 $\SU{2}$ subgroups of $\SU{3}$. Then, $R_{\textrm{local}}(x')$ can be re-written as
\begin{equation}
R_{\textrm{local}}(x') = \sum_{i,j=1}^{4} \frac{1}{2}\,\Omega_{i}\,a_{ij}\,\Omega_{j} - \sum_{i}^{4}\Omega_{i}\,b_{i} + c.
\end{equation}
Where $a_{ij}$ are elements of a real, symmetric matrix, $b_{i}$ a real vector and $c$ a real constant, all dependent only on $U_{\mu}(x')$ and $U_{\mu}(x' - \hat{\mu})$. A list of these coefficients can be found in the appendix. The global maximum of this quantity can be found using, e.g., the method of Ref.~\cite{QCQPpaper}. We then update the relevant links with the $\SU{3}$ gauge transformation $\Omega(x')$, and iterate over lattice sites. Each iteration over $\SU{2}$ subgroups and then lattice sites constitutes a single sweep. We perform 20,000 such sweeps on each configuration, after which the maximum relative change in $R_{\textrm{local}}(x)$ over a single sweep was found to be less than $10^{-10}$. \par
\begin{figure}[t!]
\includegraphics[width=\columnwidth]{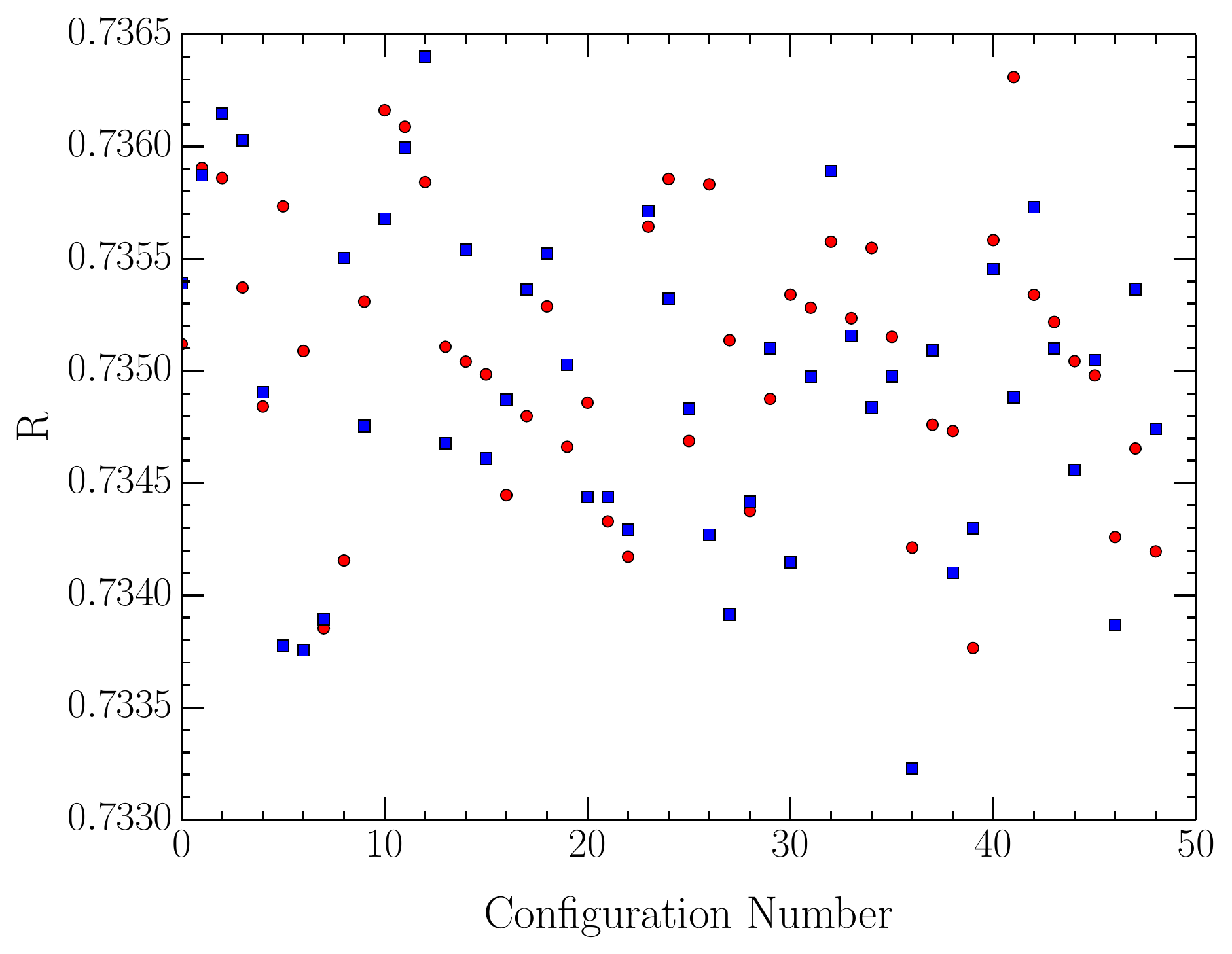}
\caption{Values of the maximal centre gauge fixing functional, $R$, after gauge fixing on a configuration-by-configuration basis. Square points in blue provide values from the original ensemble generated in the Markov chain. Circular points in red correspond to values found after performing a random gauge transformation before gauge fixing.}
\label{Rcomp}
\end{figure}
We note that although we have globally maximised $R_{\textrm{local}}(x)$ for $\Omega(x)$, we have only found a local maximum of $R$. We illustrate this Gribov copy issue in Fig.~\ref{Rcomp}; we show the values of $R$ obtained on the same configurations fixed to maximal centre gauge starting from two distinct random gauges. Although we have reached similar values of $R$ both times, there is no correspondence on a configuration-by-configuration basis; variation is well outside of the stability of the gauge fixing algorithm, indicating we have reached distinct local maxima on each occasion. This issue means we are unable to uniquely identify the vortex matter of a configuration; some of the physical centre vortices embedded in the lattice may be missed by the vortex identification procedure. Herein, we have selected the original ensemble generated in the Markov chain, denoted by the square blue points in Fig.~\ref{Rcomp}. \par
While choices of gauge fixing functional exist which avoid the issue of Gribov copies, i.e. Laplacian centre gauge \cite{Vink:1992ys,Vink:1994rb}, the vortices found in this way do not scale appropriately in the continuum limit, and so lack an interpretation as physical objects \cite{Langfeld:2003ev}. By contrast, the vortex density of objects found through Mesonic Maximal Centre Gauge fixing have a density independent of lattice spacing, and thus admit a continuum interpretation \cite{Bowman:2010zr,Langfeld:2003ev}.\par 
After gauge fixing, we decompose links into polar form as
\begin{equation}
\trace \, U_{\mu}^{\Omega}(x) = r_{\mu}(x)\,\exp \big{[}i\theta_{\mu}(x)\big{]},
\end{equation}
and thus select the centre element $Z_{\mu}(x) = \exp\big{[}\frac{2\pi i}{3} m_{\mu}(x)\big{]}$, $m_{\mu}(x) \in \{-1,0,1\}$, such that $\frac{2\pi i}{3}m_{\mu}(x)$ is closest to $\theta_{\mu}(x)$. \par
\begin{figure}[htbp]
\includegraphics[width=\columnwidth]{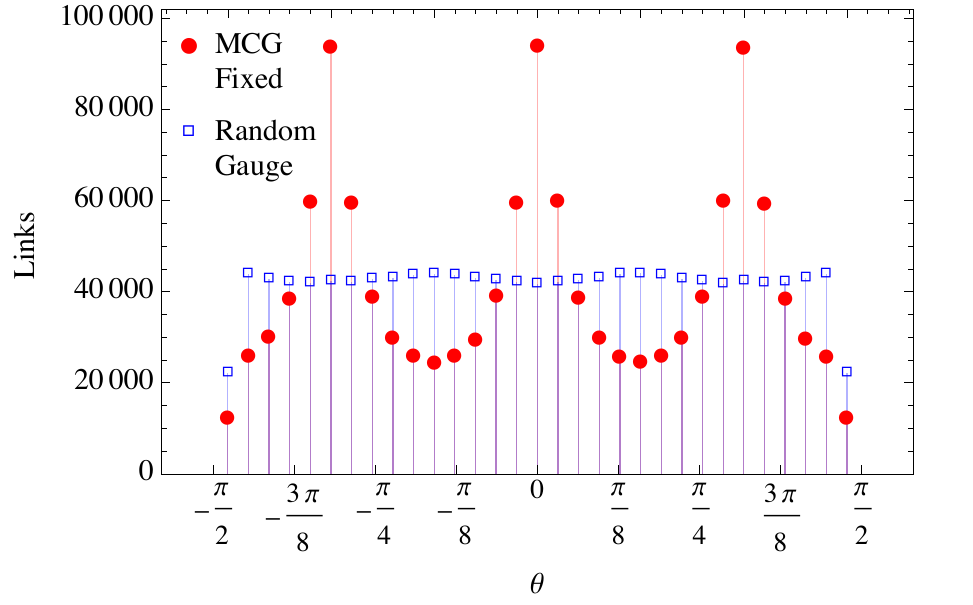}
\caption{Histogram of the phases of the trace of links on a sample configuration before and after fixing to maximal centre gauge. Values after MCG fixing are denoted by the red circles, and random gauge by blue squares.}
\label{Thetahist}
\end{figure}

 
Results are calculated on $50$ pure gauge-field configurations using the L{\"u}scher-Weisz $\mathcal{O}(a^{2})$ mean-field improved action \cite{Luscher:1984xn}, with a $20^{3} \times 40$ volume at a lattice spacing of $0.125 \, \mathrm{fm}$. In Fig.~\ref{Thetahist} we show a histogram of the values of $\theta_{\mu}(x)$ found after fixing to maximal centre gauge, showing three clear peaks, corresponding to the three centre phases. \par
We thus define the three ensembles which will serve as the basis for this work:
\begin{itemize}
\item
the untouched configurations,
\begin{equation}
U_{\mu}(x);
\end{equation}
\item
the vortex-only configurations,
\begin{equation}
Z_{\mu}(x);
\end{equation}
\item
and the vortex-removed configurations,
\begin{equation}
R_{\mu}(x) = Z_{\mu}^{\dagger}(x)\,U_{\mu}^{\Omega}(x).
\end{equation}
\end{itemize}
In order to remove potential bias from the vortex-only configurations being in the gauge where all elements are diagonal, we perform a random gauge transformation on them before smoothing. 

\section{Smoothing}
\label{sec:smoothing}

Motivated by the recent results of Ref.~\cite{Trewartha:2015nna}, showing that vortex-only configurations are able to reproduce dynamical mass generation after 10 sweeps of cooling, we perform smoothing on our ensembles. We seek to quantify the extent to which the underlying vacuum structure of vortex-only and untouched configurations is similar after smoothing.\par
Smoothing is known to reveal a gauge field background consisting of a ``liquid'' of (anti-)instanton-like objects \cite{Teper:1985rb,Ilgenfritz:1985dz}. While instanton/anti-instanton pairs continue to annihilate under smoothing, isolated instanton like objects remain stable under further smoothing, as long as over-improved or highly improved actions are used. (Anti-)Instantons are associated with approximate zero-modes of the Dirac operator \cite{Ilgenfritz:2008ia}, and thus through the Banks-Casher relation \cite{Banks:1979yr} contribute to a non-zero quark condensate, an order parameter for the dynamical breaking of chiral symmetry. \par
We will thus focus in particular on the instanton degrees of freedom of our ensembles. We will smooth in two ways; cooling \cite{Bonnet:2000dc,Bonnet:2001rc,Bonnet:2000db,BilsonThompson:2002jk,BilsonThompson:2004ez}  using an $\mathcal{O}(a^4)$-three-loop improved action \cite{Bonnet:2000db,BilsonThompson:2002jk} and over-improved stout-link(OISL) smearing \cite{Moran:2008qd,Moran:2008ra}, an improved form of stout-link smearing tuned to retain instanton-like objects. Topological charge density is calculated using an $\mathcal{O}(a^4)$-five-loop improved definition of the field-strength tensor \cite{BilsonThompson:2002jk}. \par
\begin{figure*}[thpb]
\subfigure[]{
\label{coolActvsSw}
\includegraphics[width=\columnwidth]{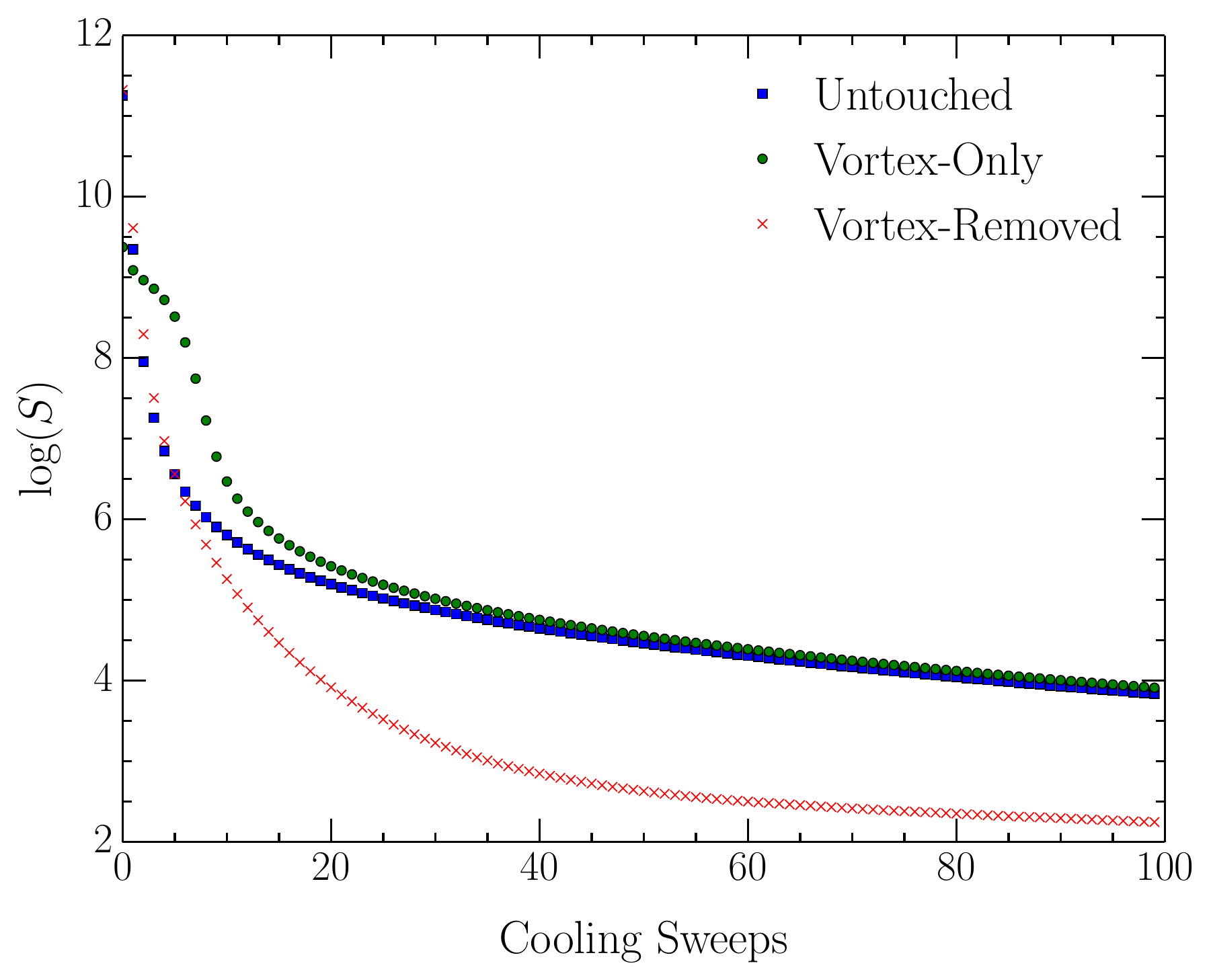}}
\subfigure[]{
\label{oislActvsSw}
\includegraphics[width=\columnwidth]{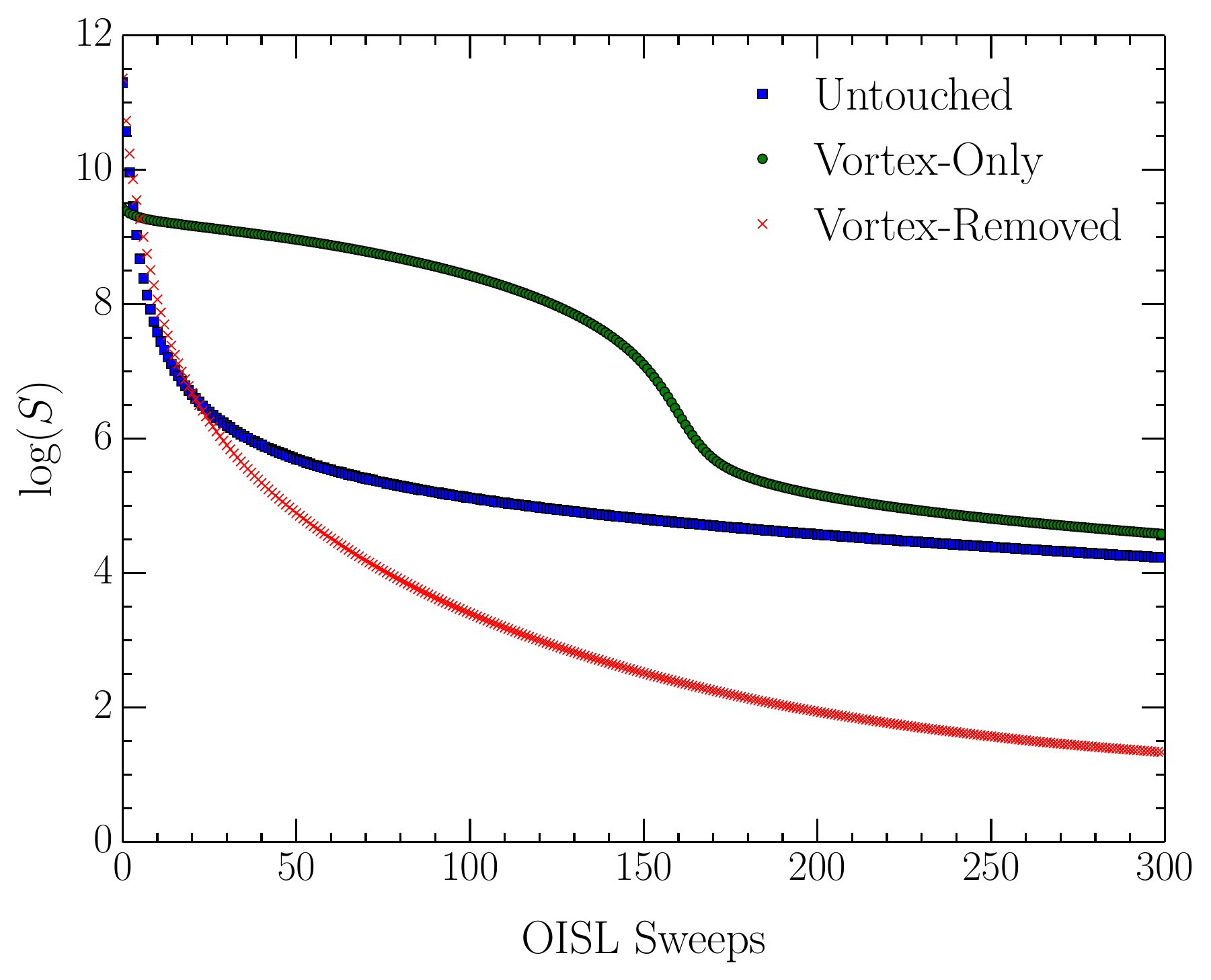}}
\caption{A log plot of the ensemble-averaged action on untouched (squares), vortex-only (circles) and vortex-removed (crosses) ensembles under cooling \subref{coolActvsSw} and over-improved stout-link smearing \subref{oislActvsSw}}.
\label{Fig:ActvsSw}
\end{figure*}
In Fig.~\ref{Fig:ActvsSw} we have plotted the ensemble-averaged action on our three ensembles as a function of smoothing sweeps. The untouched ensemble behaves as expected; the first few sweeps of smoothing rapidly removes action, as quantum fluctuations away from the classical solution are removed. Then, after about 10 sweeps of cooling or 30 of OISL smearing, a stable background of instanton-like objects is revealed, and action is very slowly removed from the lattice as these can only disappear via pair-annihilation. \par
The vortex-removed ensemble shows a very rapid loss of action, well below the untouched, and settles at a much lower value at high amounts of both cooling and OISL smearing. This corresponds to observations in Ref.~\cite{Trewartha:2015nna} that vortex removal destabilises instanton-like objects, and so after smoothing we have produced a mostly trivial, low action background. \par
The vortex-only ensemble, however, behaves markedly differently. The vortex-only configurations start very far from a classical solution to the QCD equations, and so have high action, decreasing only slowly under smoothing. Eventually, a rapid loss of action occurs. After around 20 sweeps of cooling, or 200 of OISL smearing, the rate of action loss decreases to be comparable to the untouched ensemble, and remains so up to high levels of smoothing. This corresponds to a dramatic change in the gauge field structure; it seems that this level of smoothing is required to produce a background of instanton-like objects on the vortex-only ensemble, which is then very stable under smoothing, analogous to that found in the untouched case. Just as vortex removal destabilised instanton-like objects, the centre vortex structure alone contains the essential information for these objects, which are then revealed through the smoothing procedure.\par

\begin{figure*}[thpb]
\subfigure[]{
\label{CoolQvsSw}
\includegraphics[width=\columnwidth]{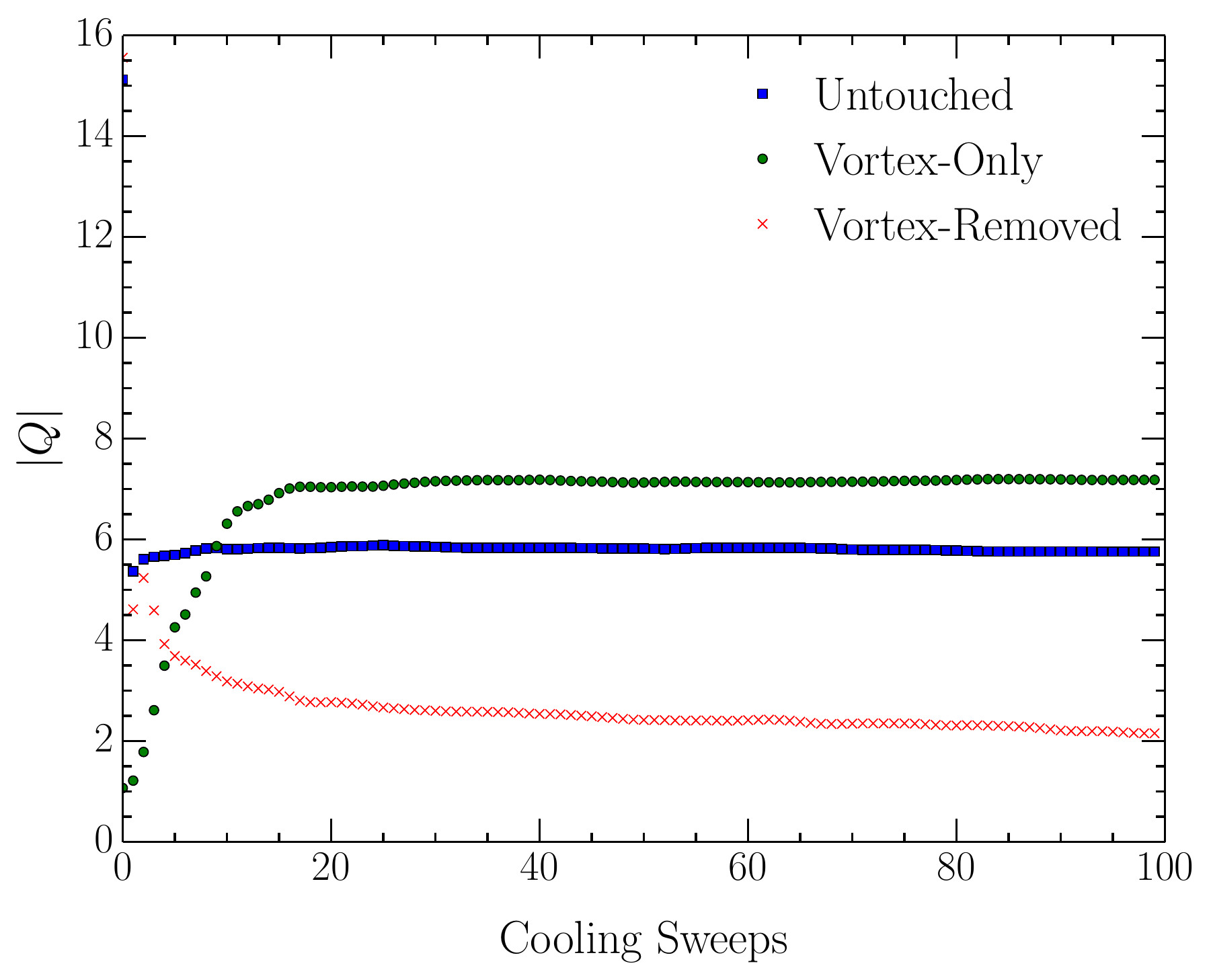}}
\subfigure[]{
\label{OISLQvsSw}
\includegraphics[width=\columnwidth]{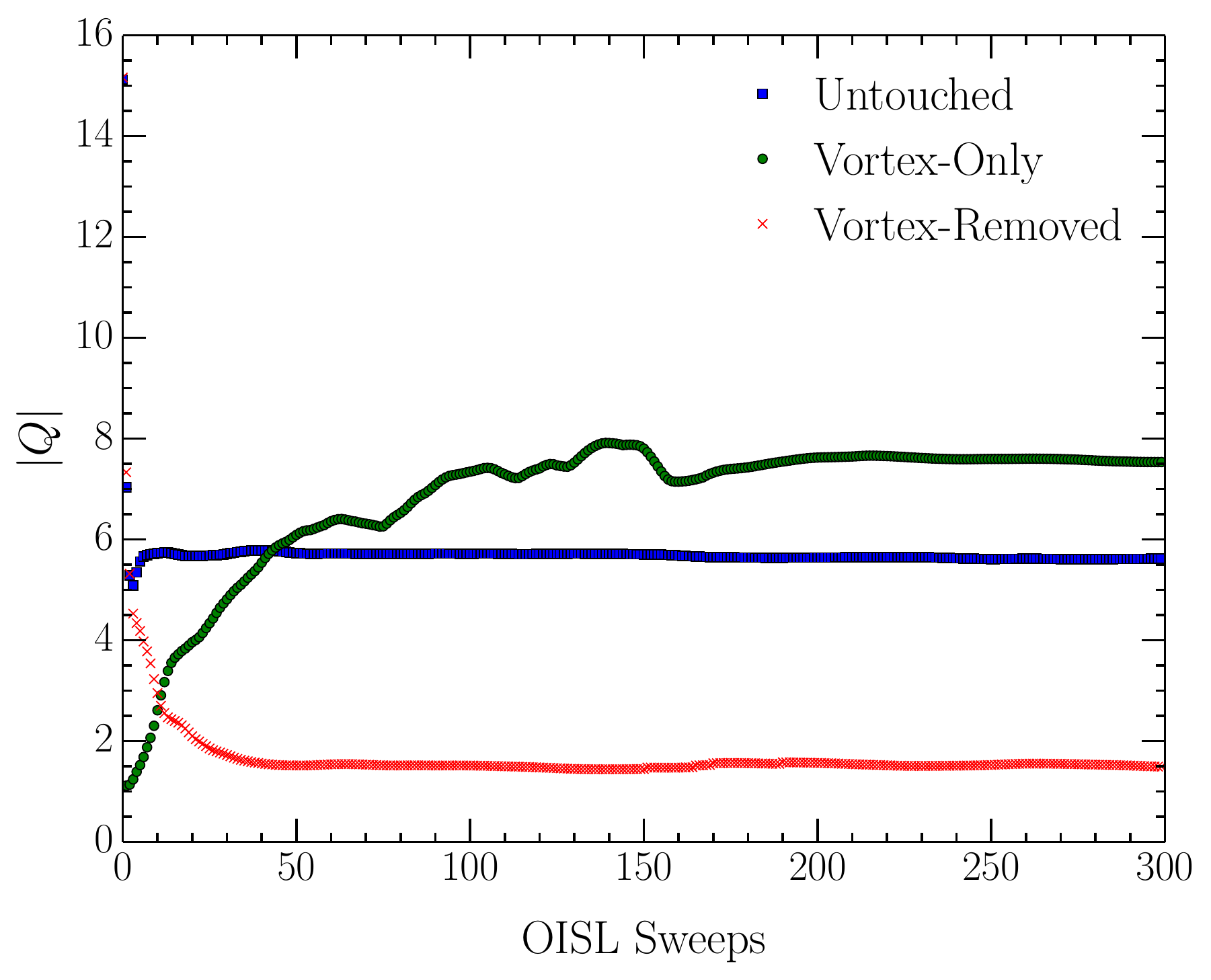}}
\caption{The ensemble-averaged absolute value of the topological charge on untouched (squares), vortex-only (circles) and vortex-removed (crosses) ensembles under cooling \subref{CoolQvsSw} and over-improved stout-link smearing. \subref{OISLQvsSw}}
\label{Fig:QvsSw}
\end{figure*}

The evolution of the ensemble-averaged absolute value of the topological charge on the three ensembles is shown in Fig.~\ref{Fig:QvsSw}. Again, the untouched ensemble behaves as expected. Initially, the integrated topological charge is poorly defined on rough configurations, but after smoothing settles on a stable value. Once our configurations have been smoothed enough to resemble an instanton liquid, instanton-anti-instanton pairs disappear from the lattice only by pair annihilation, which has no net effect on the configuration topological charge, and so this value is stable up to very high levels of smoothing. \par
Consistent with the action, the vortex-removed results show an almost empty background of very low topological charge after smoothing. The removal of centre vortices has destabilised otherwise topologically non-trivial objects, resulting in their subsequent destruction by the smoothing algorithm. The vortex-removed results do not, however, settle on no topological charge, indicating imperfections in the vortex removal procedure have left some objects intact.\par
In agreement with the action, the vortex-only results show a high degree of roughness until around 10 sweeps of cooling and 180 of OISL smearing, where they become much smoother, and display a similar result to the untouched ensemble, settling on a stable value. \par
To study the instanton degrees of freedom, we directly examine the gauge fields using the method of Ref.~\cite{Moran:2008qd}. We search the lattice for local maxima of the action density, around which we fit the classical instanton solution;
\begin{equation}
S_{0}(x) = \xi\,\frac{6}{\pi^{2}}\,\frac{\rho^{4}}{((x-x_{0})^{2}+\rho^{2})^{4}},
\end{equation}
with $\rho$ the instanton radius, and $x_{0}$ the location of the centre of the instanton, not restricted to lie on lattice sites. The scale parameter $\xi$ is introduced to ensure we fit to the shape of the action density, not the magnitude. We note that this method requires a relatively smooth background; UV noise will obscure the shape of instanton-like objects, and lead to fitting to a number of ``false positives''; local maxima of the action originating due to short-range noise, not larger-scale underlying vacuum structure. \par
\begin{figure*}[thpb]
\subfigure[]{
\label{Coolnum}
\includegraphics[width=\columnwidth]{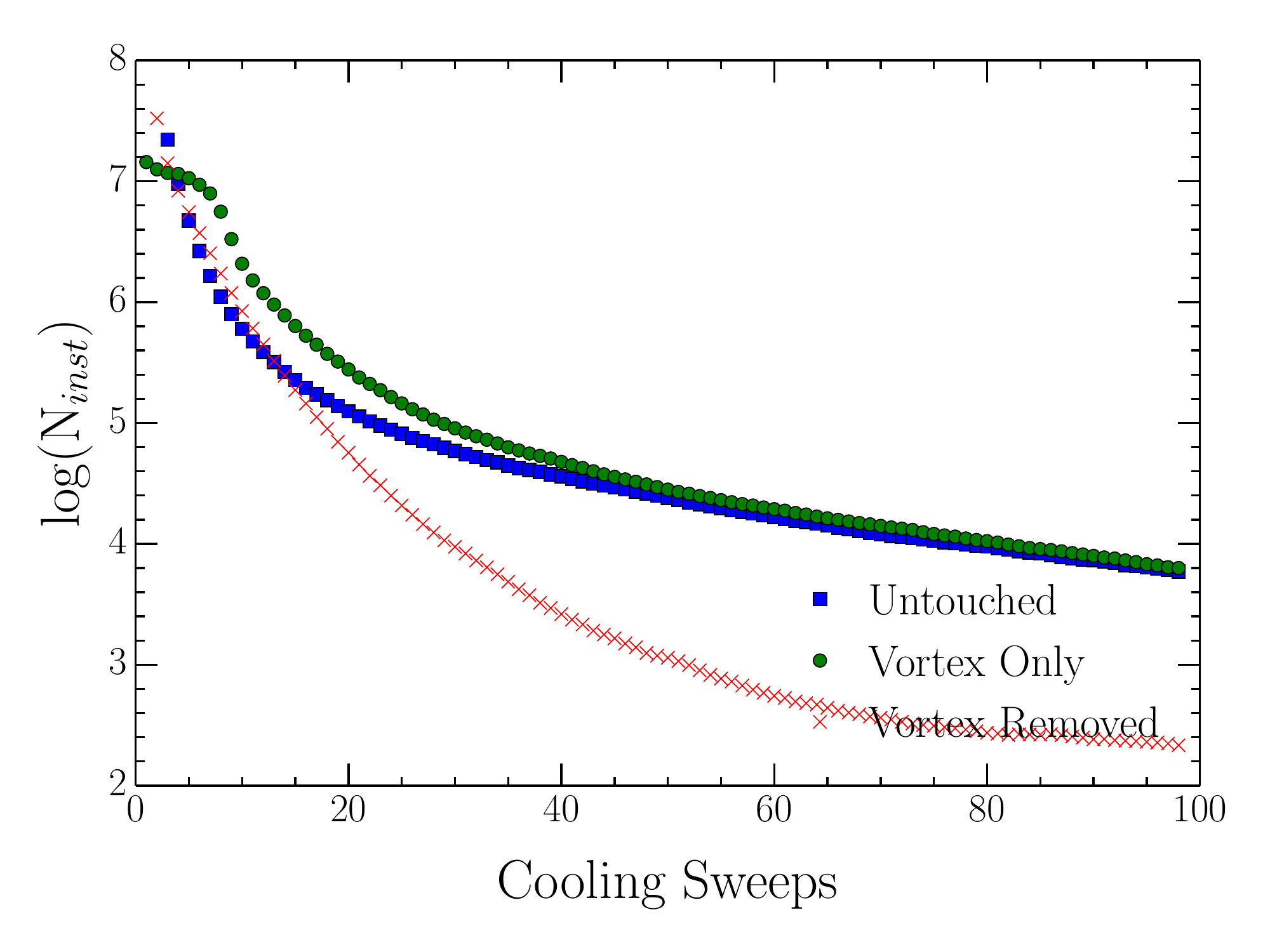}}
\subfigure[]{
\label{OISLnum}
\includegraphics[width=\columnwidth]{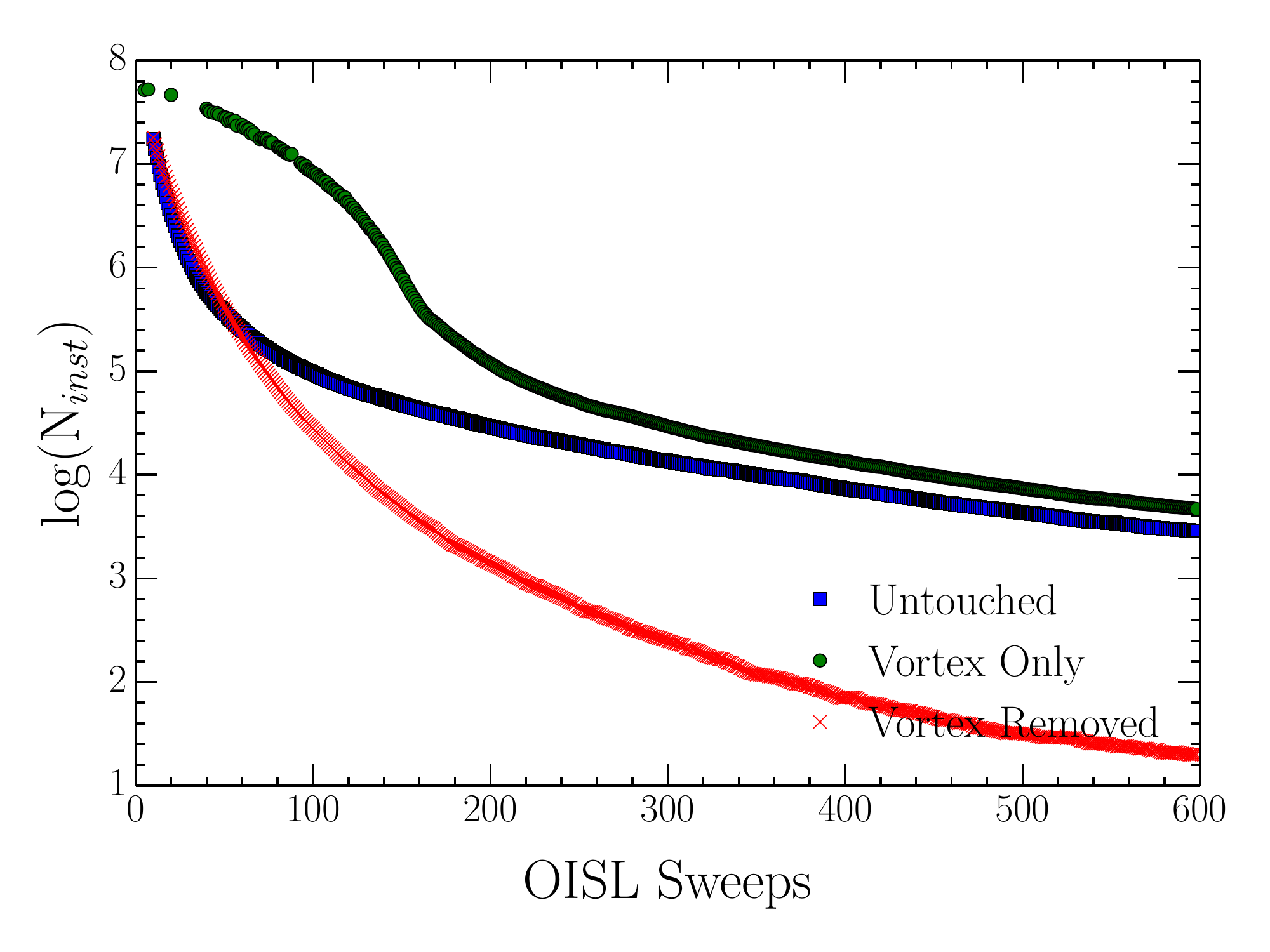}}
\caption{The ensemble-average number of local maxima of the action found on untouched (squares), vortex-only (circles) and vortex-removed (crosses) ensembles after cooling \subref{Coolnum} and over-improved stout-link smearing. \subref{OISLnum}}
\label{Fig:num}
\end{figure*}
We have plotted the ensemble-average number of local maxima of the action density as a proxy for the number of instanton-like objects on the lattice  as a function of smoothing in Fig.~\ref{Fig:num}. The rapid decrease after a small amount of smoothing in the untouched case corresponds to the appearance of a gauge-field background dominated by instanton-like objects; at small levels of smoothing local maxima are largely due to short-range noise. After this, the remaining objects are highly stable under smoothing. \par
As expected, the vortex-removed ensemble becomes almost empty under smoothing; 2 orders of magnitude fewer objects are found under high levels of smoothing. Removing centre vortices from the lattice has correspondingly destabilised instantons, leading to their destruction by smoothing algorithms, although some remain. \par
In the vortex-only case, the results seen for the integrated action and topological charge hold true for the action density; an initially noisy background, shown by a high number of local maxima of the action density, comes to resemble the instanton configuration seen on the untouched ensemble after a ``turning point'' of 10 sweeps of cooling and 180 of OISL smearing, thereafter remaining stable under smoothing. \par
\begin{figure*}[thpb]
\subfigure[]{
\label{coolrho}
\includegraphics[width=\columnwidth]{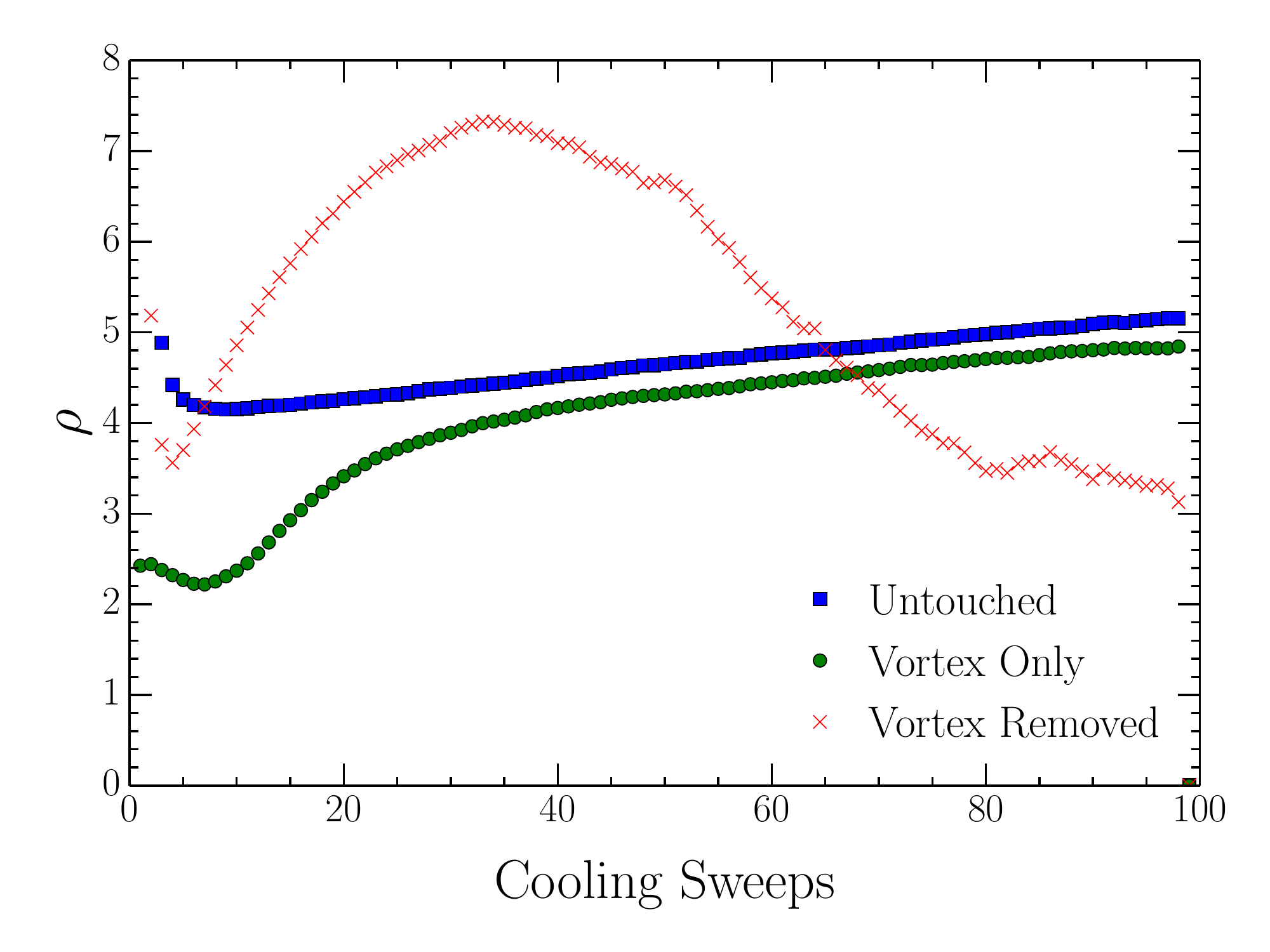}}
\subfigure[]{
\label{oislrho}
\includegraphics[width=\columnwidth]{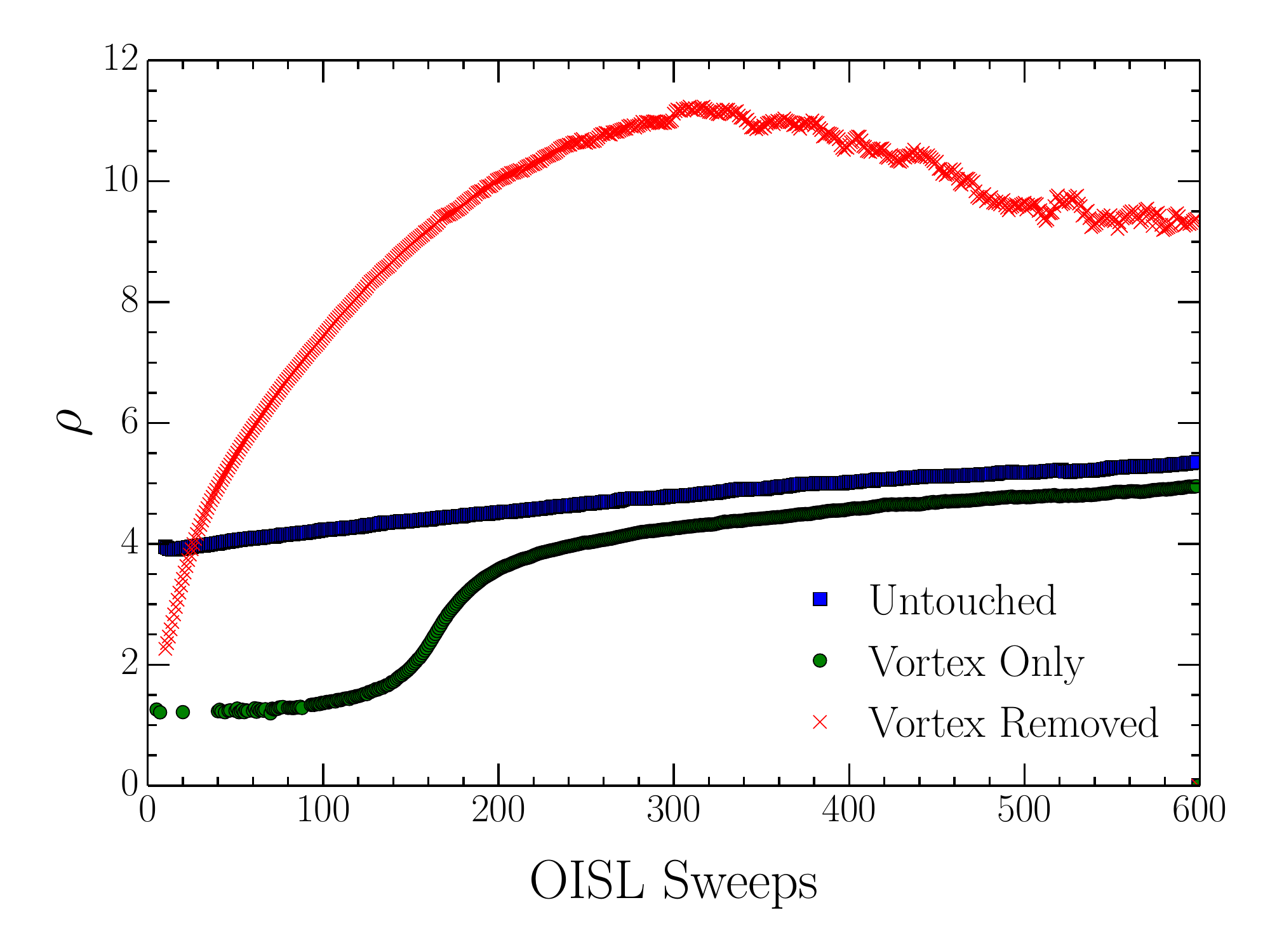}}
\caption{The ensemble-average radius ($\rho$) of instanton candidates found on untouched (squares), vortex-only (circles) and vortex-removed (crosses) ensembles after cooling \subref{coolrho} and over-improved stout-link smearing. \subref{oislrho}}
\label{Fig:rho}
\end{figure*}
In Fig.~\ref{Fig:rho} we have plotted the ensemble-average fitted radius of instanton candidates. The untouched case shows the expected shape \cite{Trewartha:2013qga}; an initial drop corresponding to the removal of false positives due to short range noise, followed by a small, steady increase as the gauge field background becomes dominated by instanton-like objects. The vortex-only result, again, shows an initially noisy background, evinced by the large number of small objects, transforming into an instanton background almost identical to that in the untouched case by around 40 sweeps of cooling or 200 of OISL smearing. The similar slopes beyond this points means a small shift in the x-coordinate would place this trajectory on top of the untouched trajectory. In the vortex-removed case, the average instanton radius becomes very large after the initial cooling. This likely corresponds to an almost empty background, where the instanton fitting algorithm has attempted to fit over large distances with an almost smooth gauge field. After a large amount of cooling, and the large objects have been smoothed away, the vortex-removed ensemble shows a small, stable number of small instantons, most likely due to Gribov copy induced inefficiencies in the vortex removal procedure failing to eliminate all centre vortices. \par
\begin{figure*}[thpb]
\subfigure[]{
\label{UTcool10}
\includegraphics[width=\columnwidth]{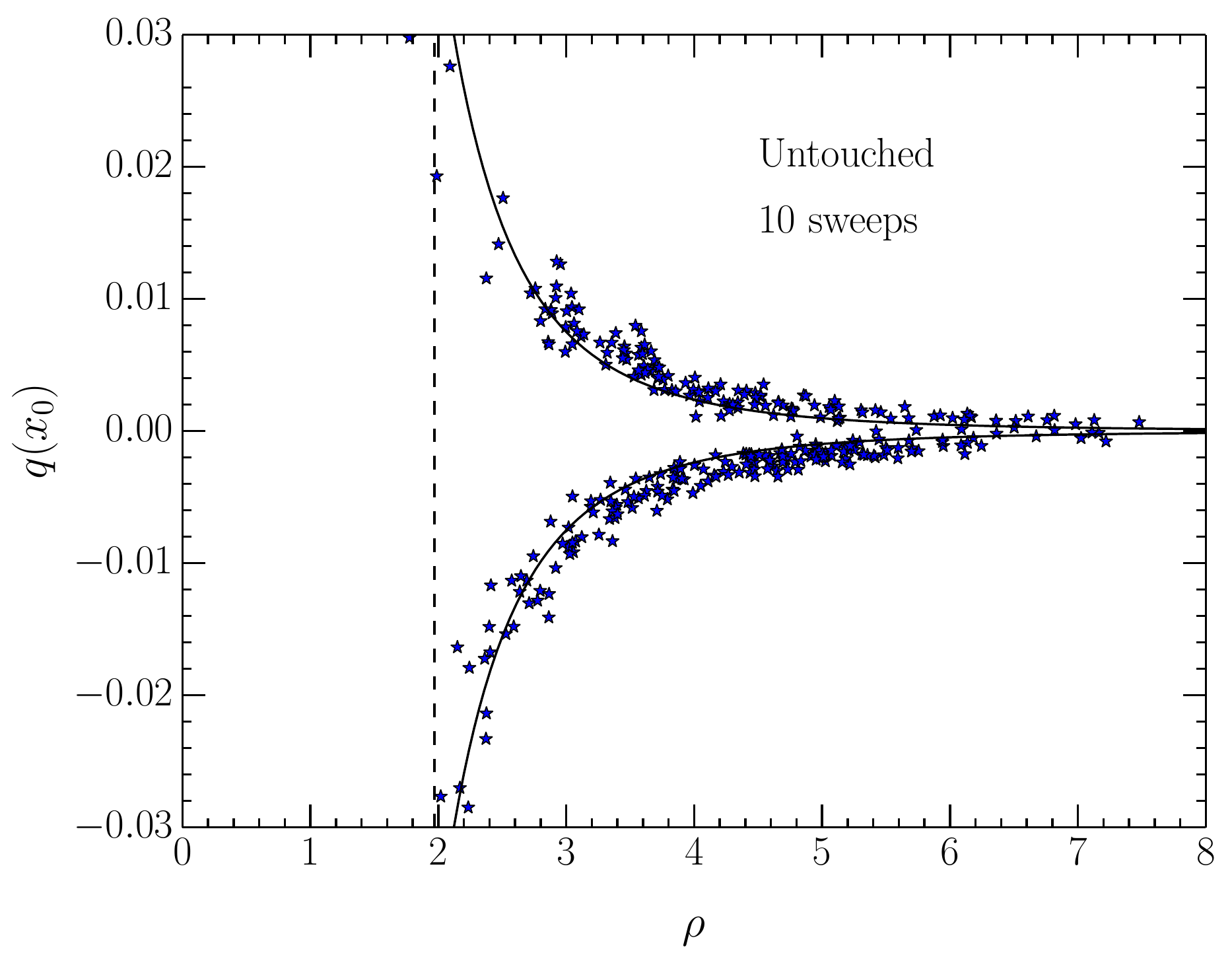}}
\subfigure[]{
\label{UTcool40}
\includegraphics[width=\columnwidth]{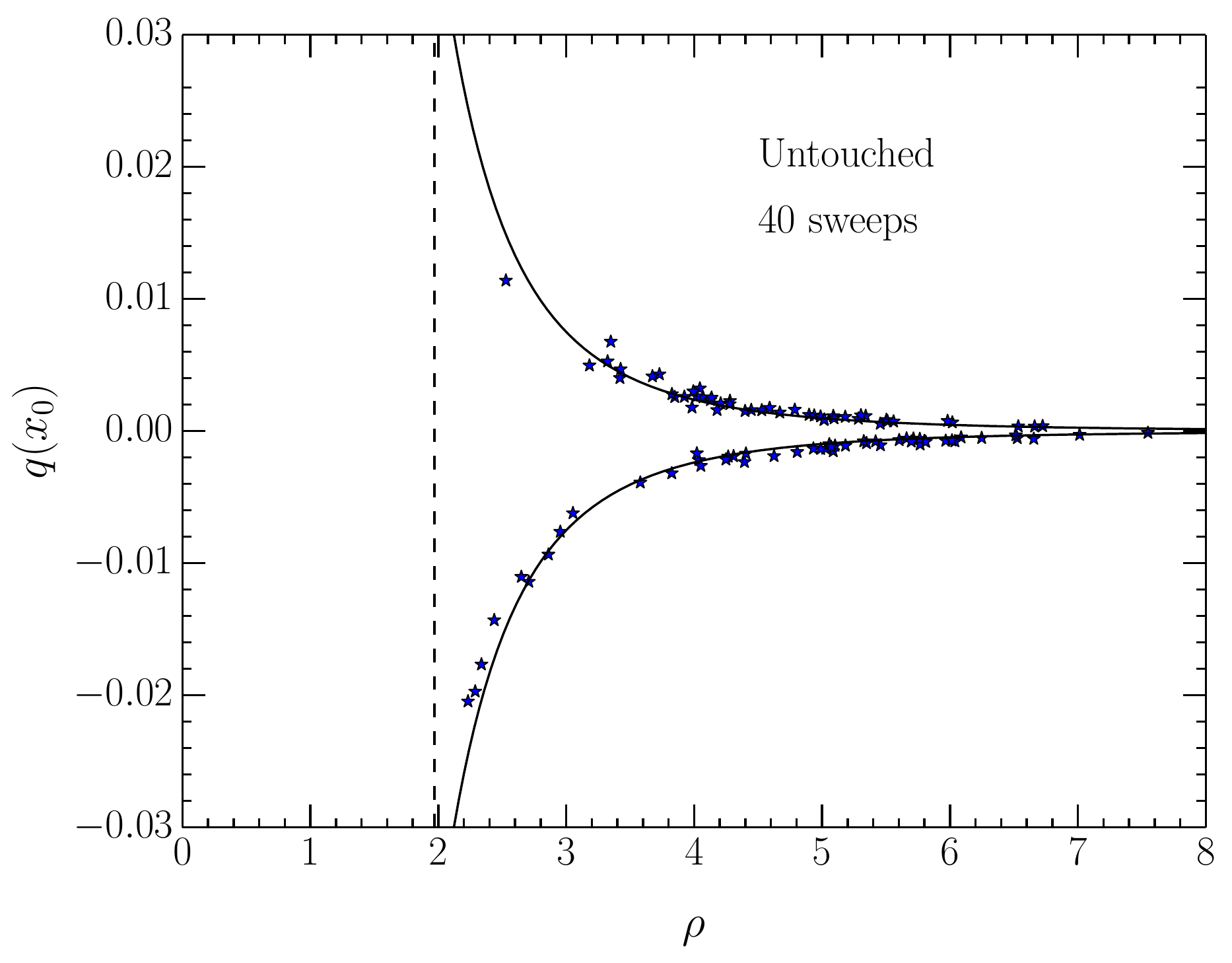}}
\subfigure[]{
\label{VOcool10}
\includegraphics[width=\columnwidth]{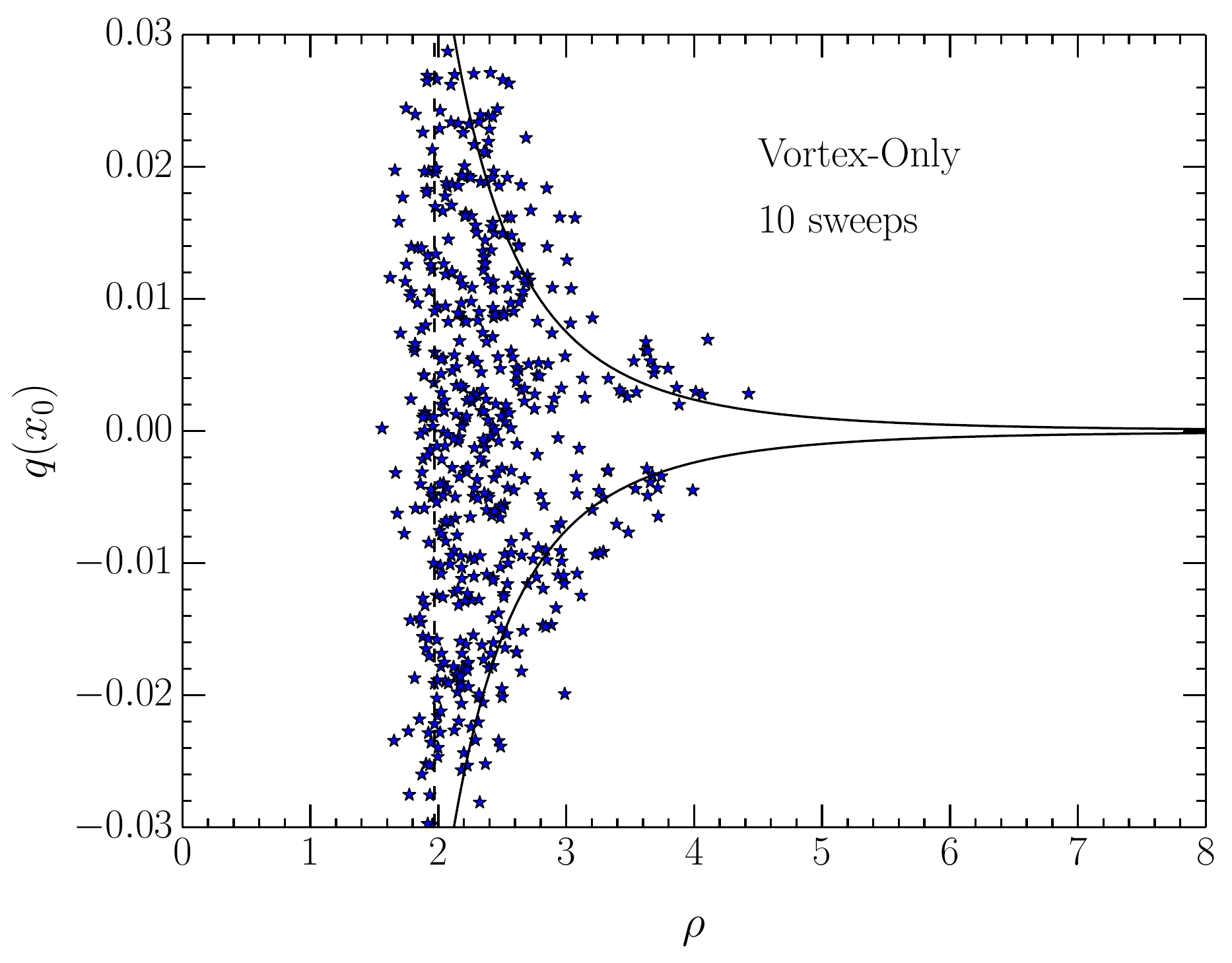}}
\subfigure[]{
\label{VOcool40}
\includegraphics[width=\columnwidth]{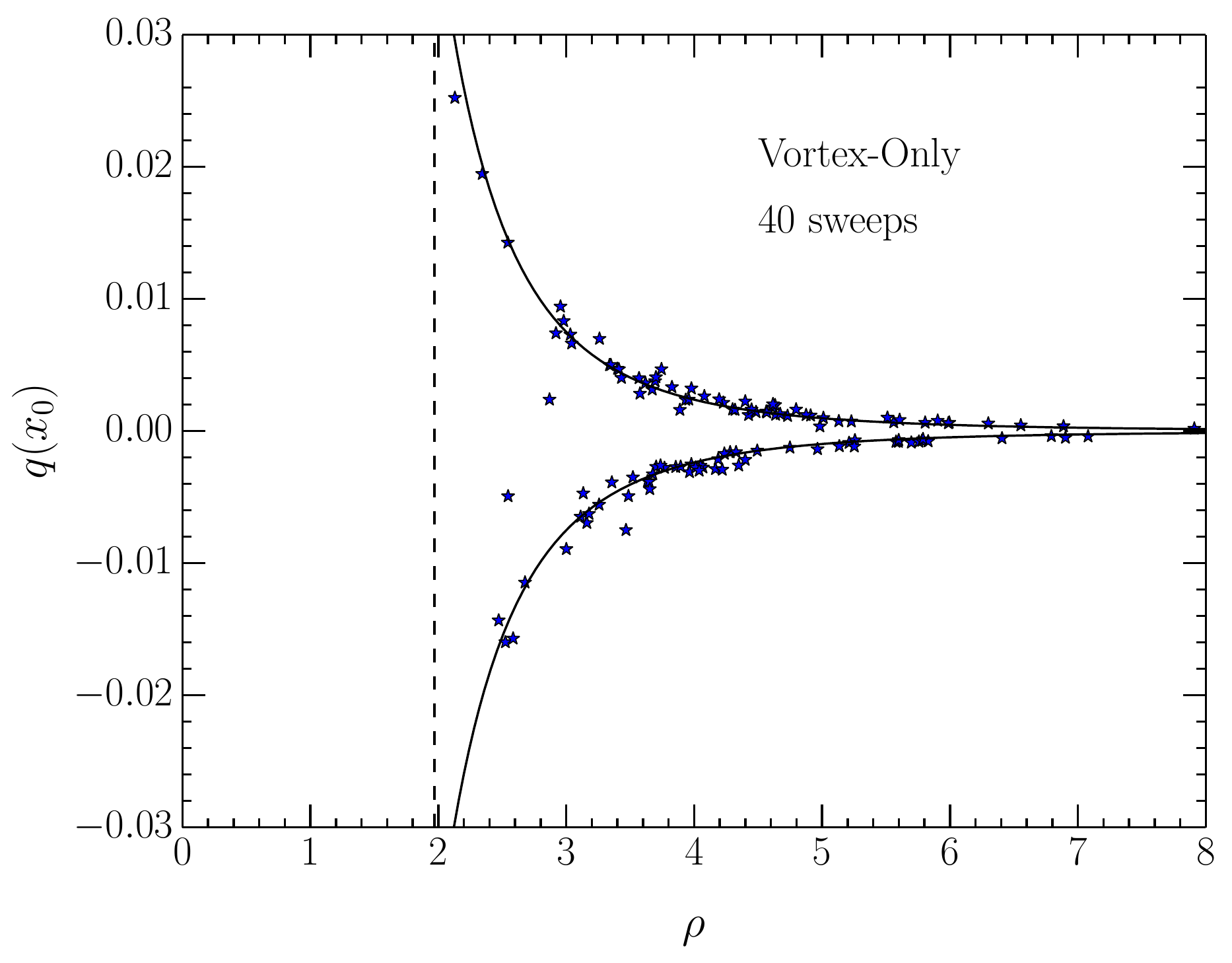}}
\subfigure[]{
\label{VRcool10}
\includegraphics[width=\columnwidth]{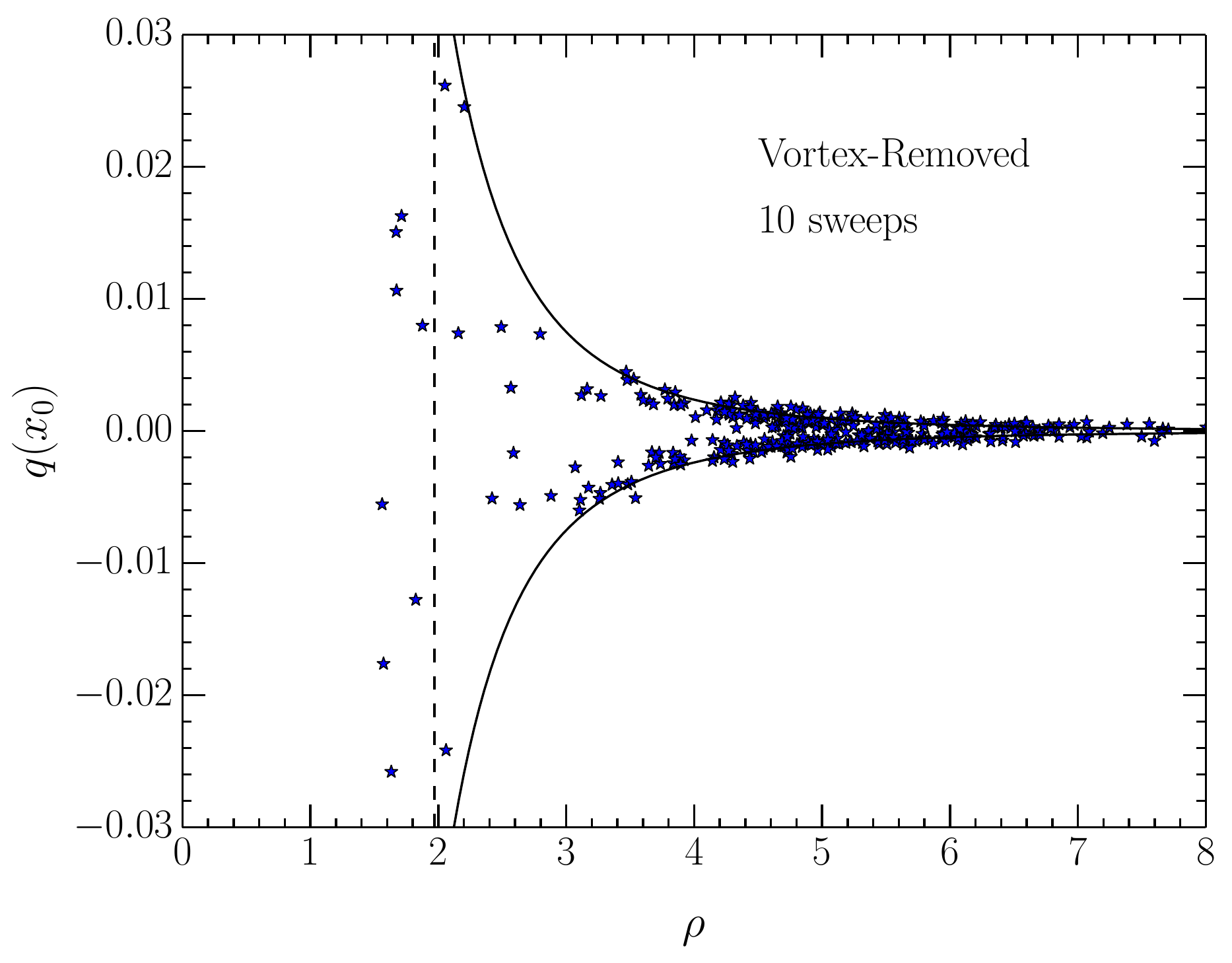}}
\subfigure[]{
\label{VRcool40}
\includegraphics[width=\columnwidth]{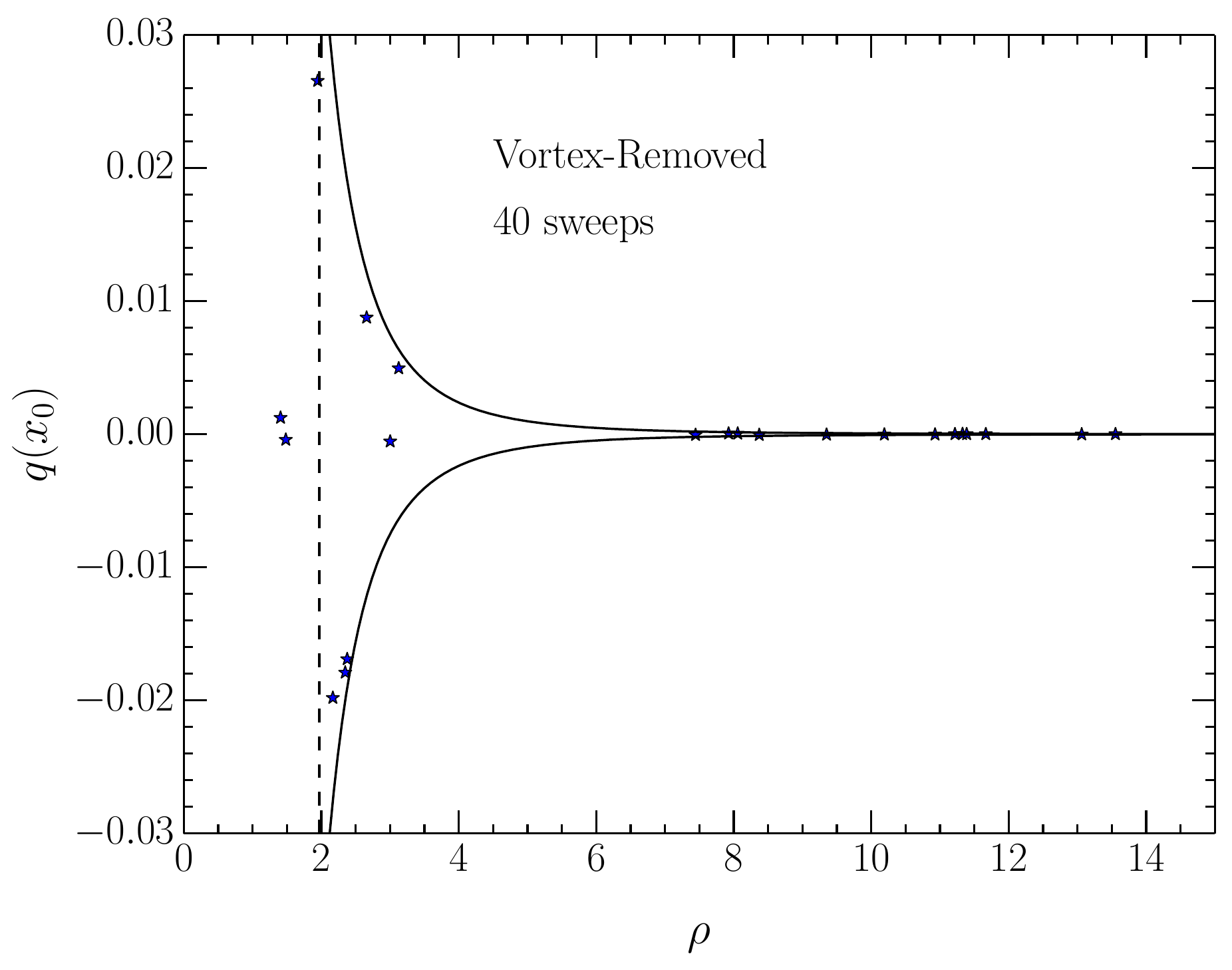}}
\caption{The values of the instanton radius, $\rho$, against the topological charge at the centre, $q(x_{0})$, on a representative configuration. Results are compared to the theoretical relationship between the instanton radius and topological charge at the centre, Eq.~(\ref{topcharge}) (solid lines), and the dislocation threshold, $1.97a$ (dashed line). Results are shown on a typical configuration for the untouched (top), vortex-only (middle), and vortex-removed (bottom) cases at 10 (left) and 40 (right) sweeps of cooling. Note the larger scale for the vortex-removed case after 40 sweeps of cooling.}
\label{Fig:cool10RvQ}
\end{figure*}

To further elucidate the situation, we compare the fitted radius ($\rho$) and measured topological charge at centre ($q(x_{0})$) of instanton candidates to the theoretical relationship,
\begin{equation}
\label{topcharge}
q(x_{0})=Q\,\frac{6}{\pi^{2} \rho^{4}},
\end{equation}
where $Q=\mp 1$ for an (anti-)instanton. This is plotted for a typical configuration in Fig.~\ref{Fig:cool10RvQ} after 10 and 40 sweeps of cooling. Other configurations examined show similar behaviour. All instantons found on the untouched configuration very closely follow the theoretical relationship after 10 sweeps of cooling, and continue to do so at 40, although their number has been greatly reduced by pair annihilation. While the vortex-only configuration after 10 sweeps of cooling has a large number of instantons lying close to the theoretical relationship, there are also a large number of small objects lying far from the relationship. After 40 sweeps, only instanton-like objects remain. It seems that while instanton-like objects are present on the vortex-only configurations after 10 sweeps, it takes far longer than in the untouched case for smoothing to remove noise and thus create an instanton dominated background. Notably, there is no direct correlation between the instanton-like objects found in the untouched and vortex-only cases. While we have recreated a similar instanton-liquid like background in the vortex-only case, we have not recreated specific individual objects. By contrast, the vortex-removed background is almost empty after 40 sweeps, with only a small number of objects remaining. The way the smoothing algorithm destroys these otherwise non-trivial objects can be seen at 10 sweeps; these objects are enlarged, resulting in a smaller topological charge at centre, until being smoothed away. This explains the observation of Fig.~\ref{Fig:rho} that at intermediate ranges the objects on vortex-removed configurations are very large, before the average size shrinks dramatically.
\par
\begin{figure*}[thpb]
\subfigure[]{
\label{UTcoolq40}
\includegraphics[width=\columnwidth,trim=0cm 7cm 0cm 0cm,clip=true]{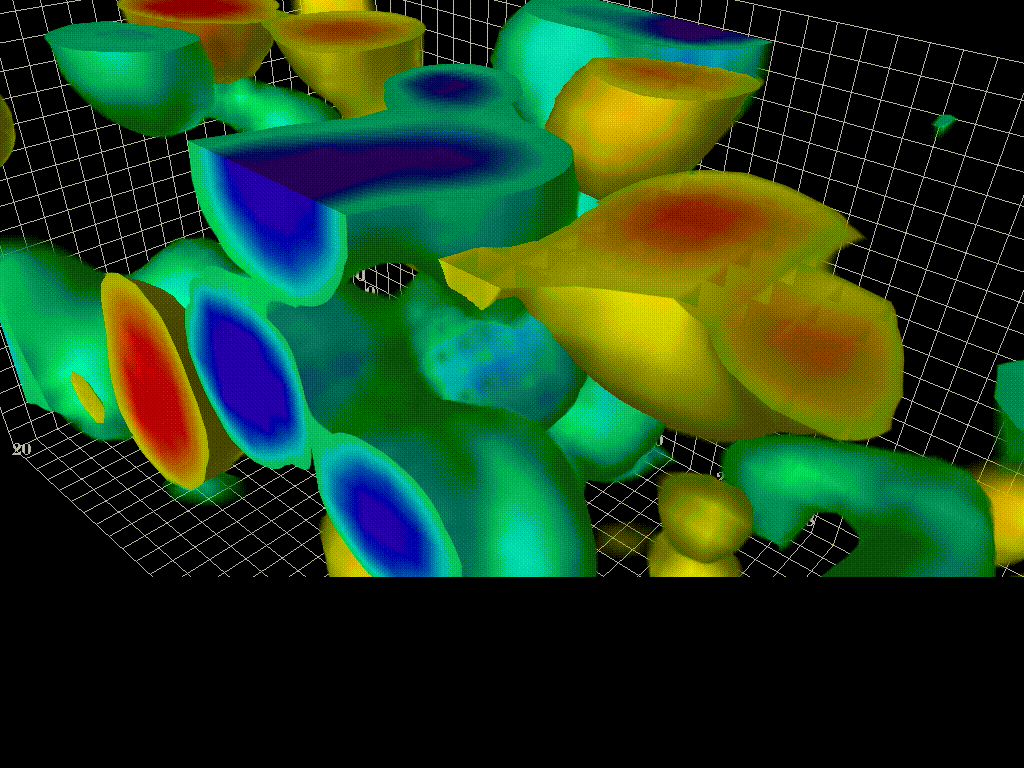}}
\subfigure[]{
\label{VOcoolq40}
\includegraphics[width=\columnwidth,trim=0cm 7cm 0cm 0cm,clip=true]{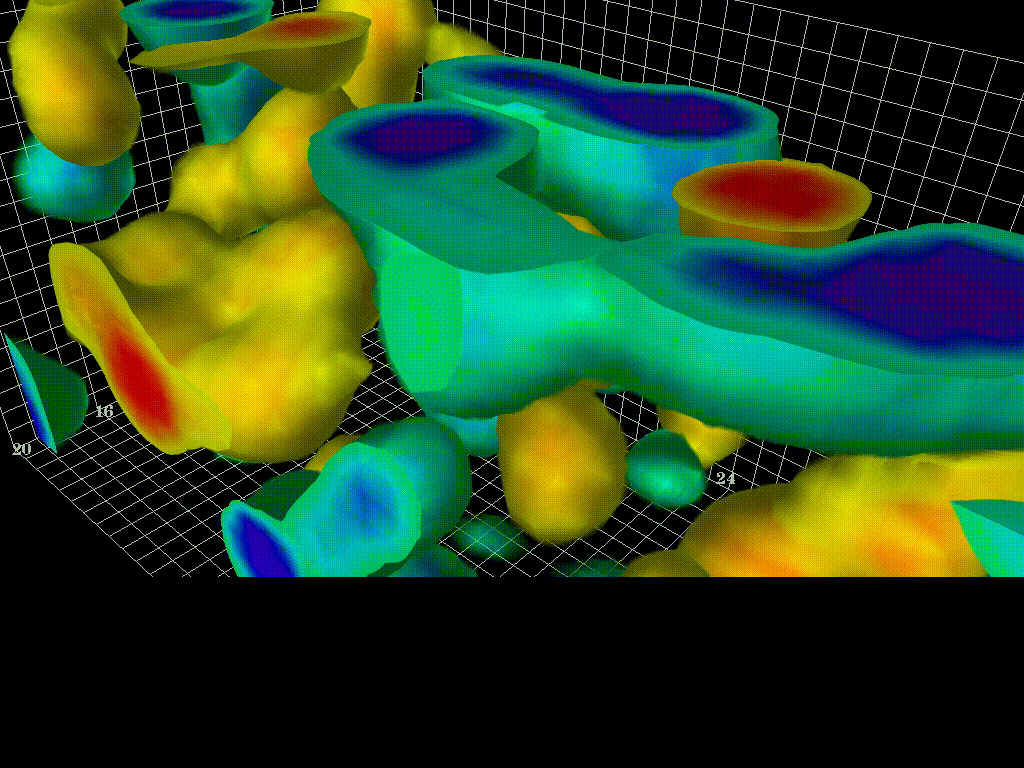}}
\caption{A visualisation of the topological charge density after 40 sweeps of cooling on a representative configuration, in both untouched \subref{UTcoolq40} and vortex-only \subref{VOcoolq40} form. Contours of positive topological charge density are plotted in yellow, and negative in blue, for a level set.}
\label{Fig:cool40q}
\end{figure*}

We have illustrated the qualitatively similar vortex-only and untouched backgrounds through the topological charge density in Fig.~\ref{Fig:cool40q}. We have plotted a level set of the topological charge density after 40 sweeps of cooling on a single configuration, both in its untouched and vortex only states. In both cases instanton-like objects are apparent, appearing as regions of high topological charge density. There is no 1-1 correspondence, however, between the objects found on the untouched configuration and on the vortex-only one.

\par 
The mechanism by which dynamical chiral symmetry breaking is apparent on vortex-only configurations after 10 sweeps of cooling is thus made clear; although there is still a large amount of noise present, instanton-like objects have begun to appear. Finding instanton-like objects through direct examination of the gauge field requires a largely smooth background, while their presence can be felt through dynamical chiral symmetry breaking manifest in the overlap quark propagator much earlier.
\par
Combined, the topological charge, action, and instanton characteristics paint a unified picture; under vortex removal, otherwise topologically non-trivial objects are destabilised, resulting in a trivial background under smoothing. By contrast, the vortex-only results, despite consisting solely of the centre degree of freedom of the original ensembles, are able to create the same long-range gauge field characteristics as seen in the untouched case after smoothing. The centre vortices, it seems, contain the ``seeds'' of instanton-like objects - the information necessary to recreate the topological structure of the gauge field background. Notably, both methods of smoothing have produced similar behaviour; the centre vortex degree of freedom robustly contains all information about the long-range vacuum structure, regardless of the smoothing method used to expose it. The measures of vacuum structure we have considered are also in agreement as to the amount of smoothing required for this to happen; it seems that at around 10 sweeps of cooling or 180 of OISL smearing, instanton-like objects have begun to appear on vortex-only backgrounds, and that by around 40 sweeps of cooling or 200 of OISL smearing the resulting vacuum structure closely resembles that of the untouched on a macroscopic level, after the same amount of smoothing. \par
Having established the similarity of results under cooling and OISL, we will henceforth restrict calculations to cooled ensembles.

\section{Static Quark Potential}
\label{sec:SQP}

Having seen the creation of a gauge field background on vortex-only configurations similar to the untouched case after smoothing, we now turn our attention to the ability of this background to recreate confinement. It has long been known \cite{Langfeld:2003ev,Stack:2002sy} that while vortex-removal results in the removal of the string tension, and thus confinement, vortex-only ensembles, without smoothing, can reproduce only approximately $66\%$ of the string tension.\par
\begin{figure}[thpb]
\includegraphics[width=\columnwidth]{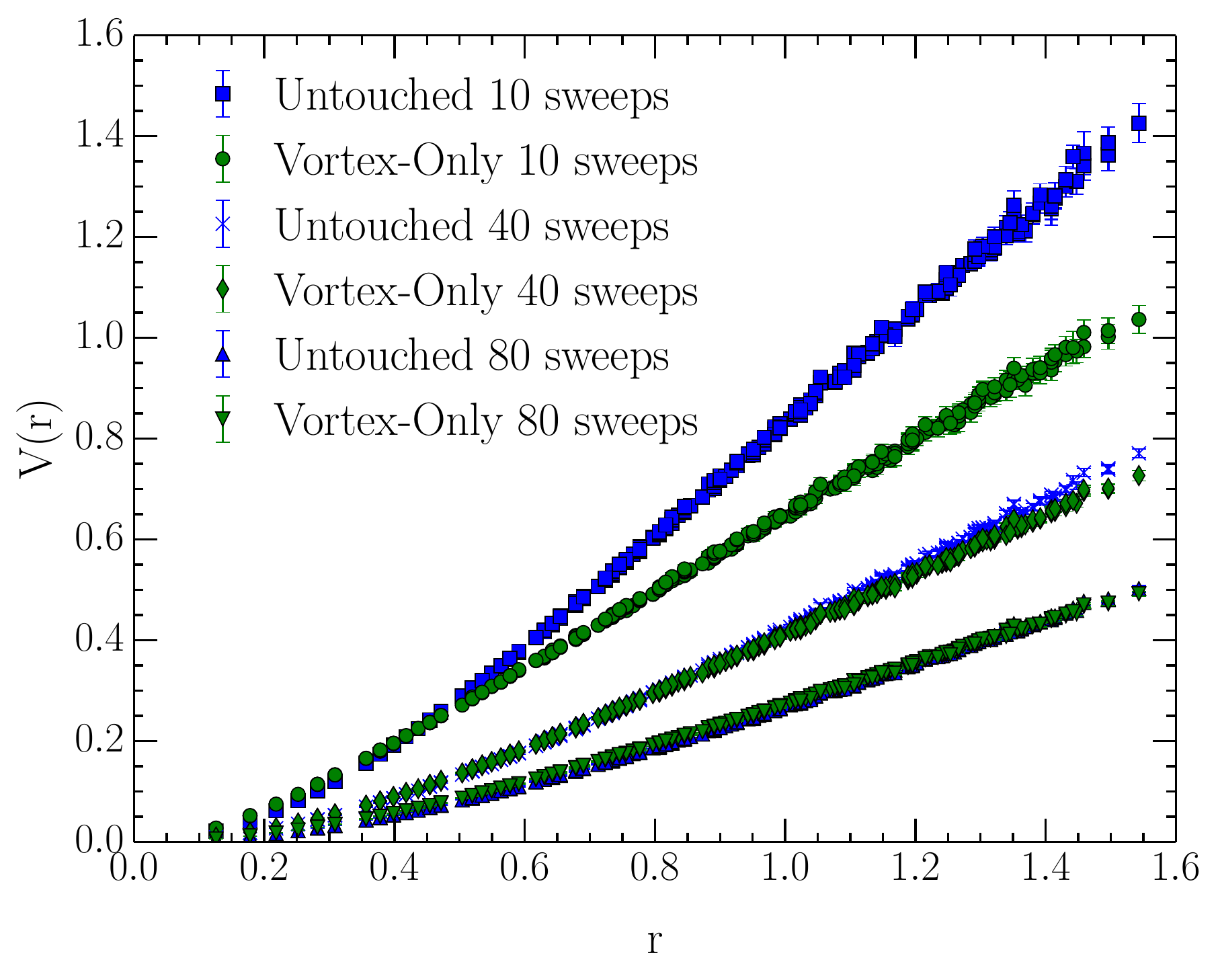}
\caption{The static quark potential on untouched (blue) and vortex-only (green) configurations after 10, 40, and 80 sweeps of cooling. Note that the 80 sweeps untouched potential is hidden behind the 80 sweeps vortex-only potential.}
\label{Fig:SQPc}
\end{figure}
\begin{table}[b]
\caption{Value of the string tension (lattice units) found on untouched (UT) and vortex-only (VO) ensembles after 10, 40, and 80 sweeps of cooling.}
\begin{tabular}{ ccc }
Cooling Sweeps & $\sigma_{\mathrm{UT}}$ & $\sigma_{\mathrm{VO}}$ \\
\hline
10 & 1.120(4) & 0.751(5) \\
40 & 0.641(2) & 0.592(2) \\
80 & 0.423(1) & 0.409(1) \\
\end{tabular}
\label{Tab:SQPc}
\end{table}
We thus examine the static quark potential on untouched and vortex-only ensembles under cooling, plotted in Fig.~\ref{Fig:SQPc}. At long distances, we have fit to the form $V(R) = \sigma R + c$, with fit parameter $\sigma$ summarised in table~\ref{Tab:SQPc}. At 10 sweeps of cooling, results at long range are consistent with the unsmoothed case; the vortex-only string tension is only $67\%$ of the untouched. However, after 40 sweeps of cooling, this picture has changed; the vortex-only ensemble now reproduces $93\%$ of the string tension found in the untouched case. This is accompanied by a decrease in the string tension found in the untouched case, to approximately half its initial value. By 80 sweeps, the percentage of the untouched string tension reproduced by the vortex-only ensembles has risen to $97\%$. It seems that after smoothing the degrees of freedom responsible for heavy quark confinement are also identical on the untouched and vortex-only ensembles.

We note that while 10 sweeps of cooling are sufficient to recreate dynamical chiral symmetry breaking on the vortex-only configurations, 40 are required to recreate confinement. This difference in scales suggests an increased sensitivity to roughness when examining confinement.

\section{Landau-gauge Overlap Quark Propagator}
\label{sec:QProp}

We now examine dynamical mass generation, and corresponding dynamical chiral symmetry breaking, on our ensembles, via the Landau-gauge overlap quark propagator. The Landau-gauge quark propagator provides a clear signal of the presence or absence of dynamical mass generation; the infrared enhancement (or lack thereof) of the Dirac scalar part of the propagator. We decompose the Landau-gauge quark propagator in momentum space into Dirac scalar and vector components as \cite{Bonnet:2002ih,Kamleh:2004aw}
\begin{equation}
S(p) = \frac{Z(p)}{iq\!\!\!/\ + M(p)}\, ,
\end{equation}
where $q$ is the kinematic lattice momentum \cite{Bonnet:2002ih}. $M(p)$ is the non-perturbative mass function, and $Z(p)$ contains the renormalisation information. We use the overlap fermion action, as it satisfies the Ginsparg-Wilson relation \cite{Ginsparg:1981bj}, and thus provides a lattice-realisation of chiral symmetry. This superior sensitivity to chiral effects is key to examining the effects of gauge field topology.\par 
We use the fat-link irrelevant clover (FLIC) fermion operator \cite{Zanotti:2001yb,Kamleh:2004aw,Kamleh:2004xk,Kamleh:2001ff} as the overlap kernel $D(-m_{w})$, with regulator parameter $m_{w} = 1$. We have considered three values of the overlap mass parameter, $\mu = 0.004$, $\mu = 0.012$, and $\mu = 0.022$, corresponding to physical bare quark masses of $12$ MeV, $40$ MeV, and $70$ MeV respectively. We rotate to Landau gauge using a Fourier transform accelerated algorithm \cite{Davies:1987vs}, fixing to the $\mathcal{O}(a^2)$ improved gauge-fixing functional \cite{Bonnet:1999mj}. A cylinder cut \cite{Leinweber:1998im} is performed on propagator data, and $Z(p)$ is renormalised to be $1$ at the highest momentum considered, $p \simeq 5.2$ GeV. \par
\renewcommand\thesubfigure{\alph{subfigure}}
\begin{figure*}[thpb]
\subfigure[]{
\label{M00400UTc40VOc40}
\includegraphics[width=\columnwidth]{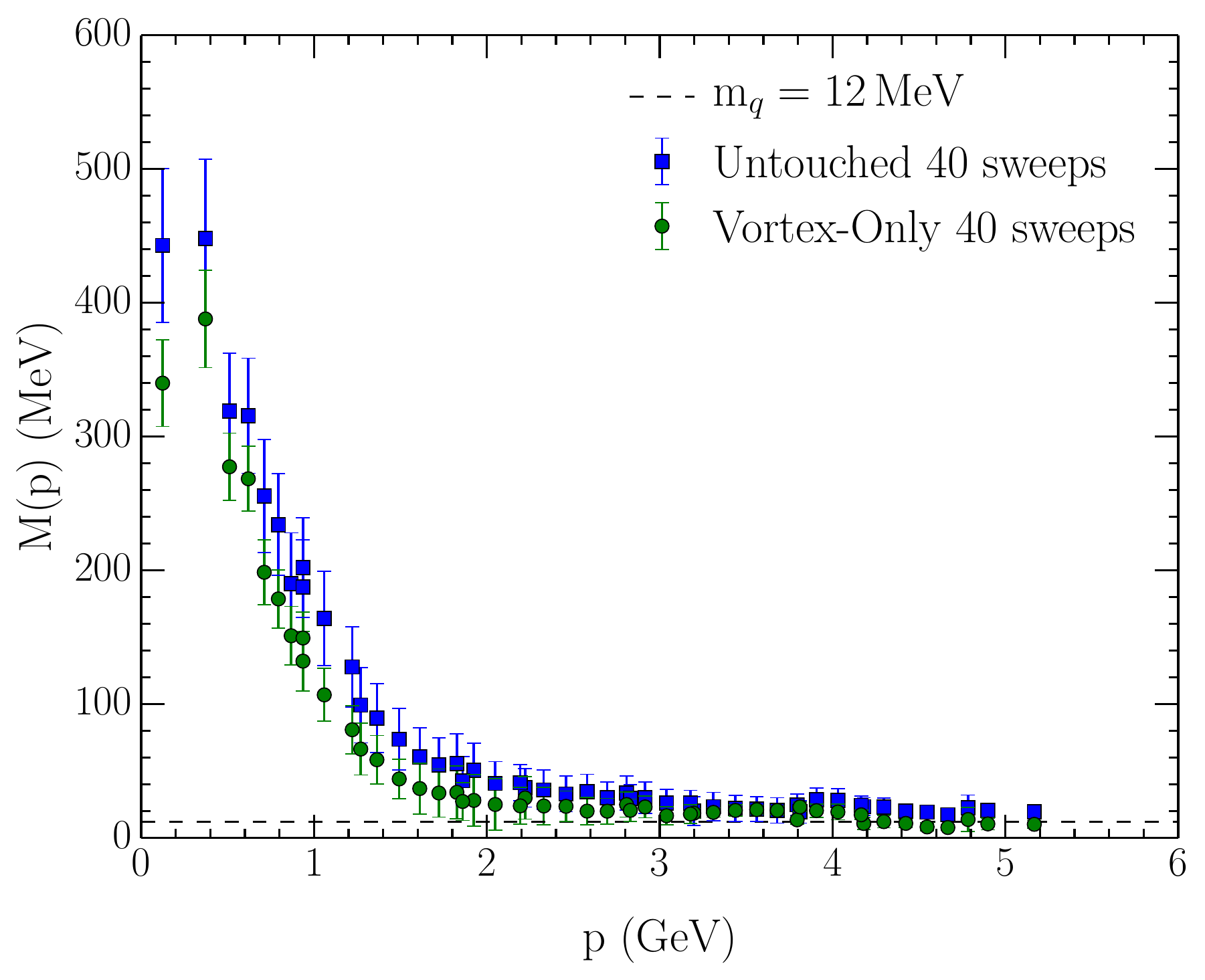}}
\subfigure[]{
\label{Z00400UTc40VOc40}
\includegraphics[width=\columnwidth]{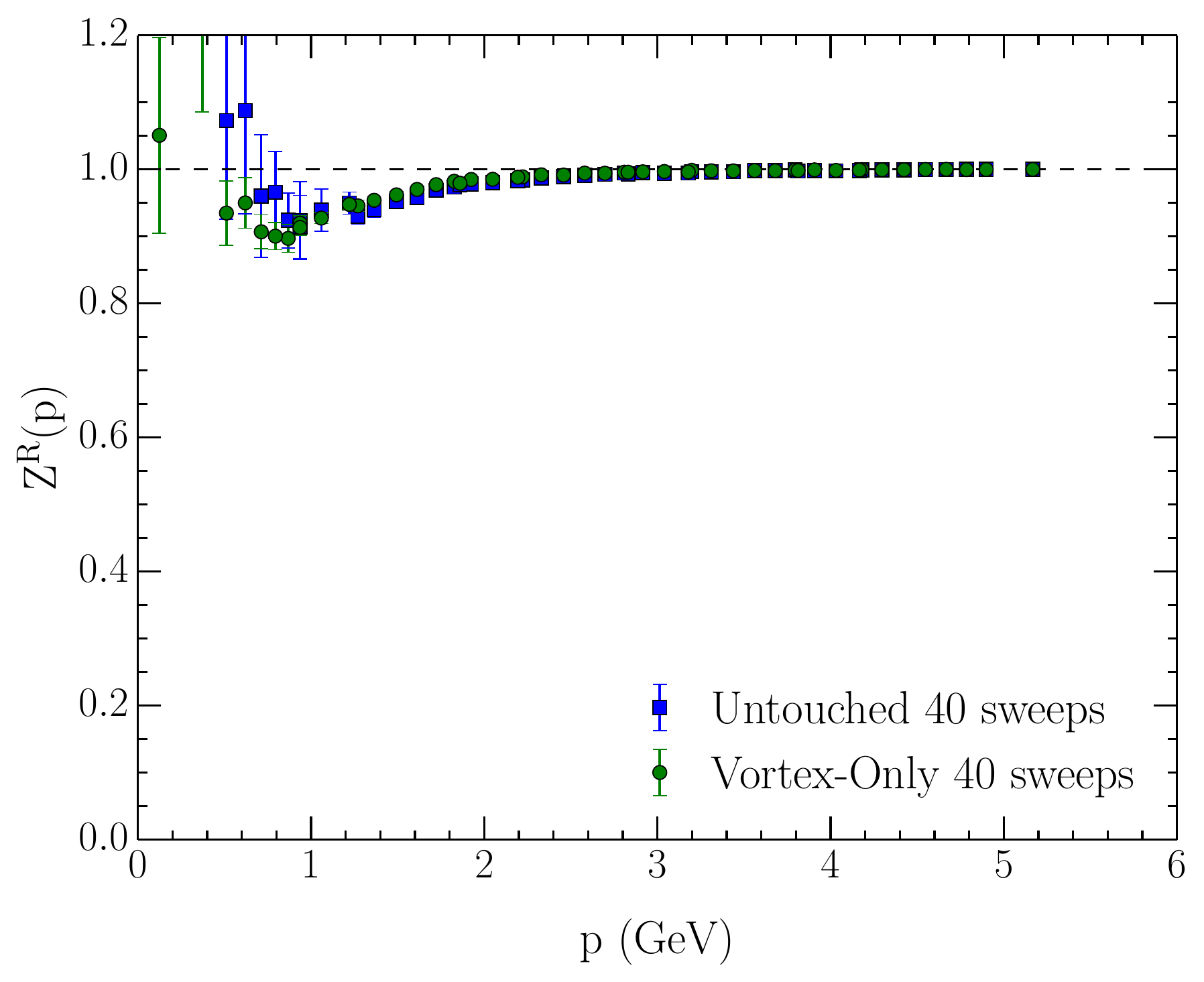}}
\subfigure[]{
\label{M01200UTc40VOc40}
\includegraphics[width=\columnwidth]{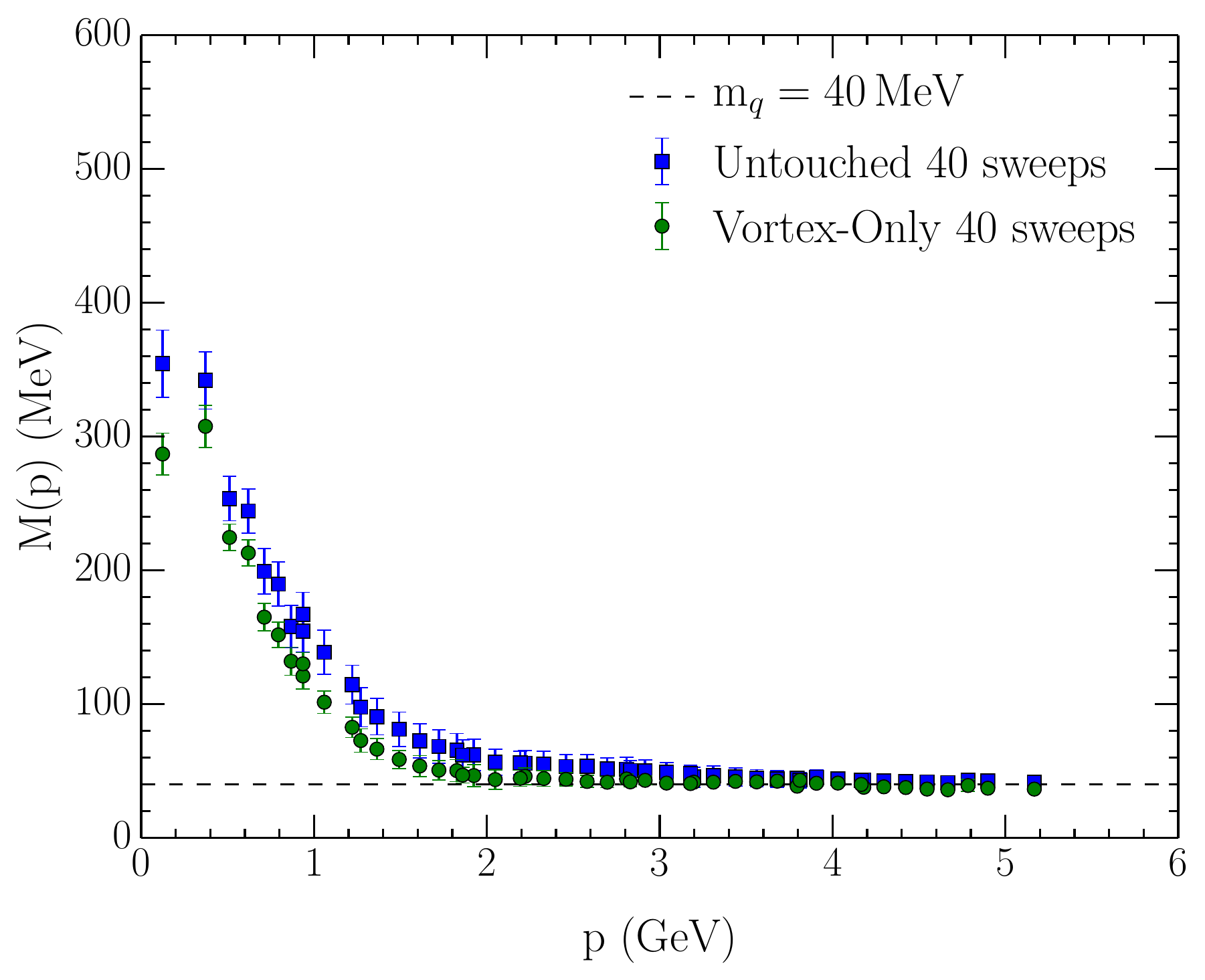}}
\subfigure[]{
\label{Z01200UTc40VOc40}
\includegraphics[width=\columnwidth]{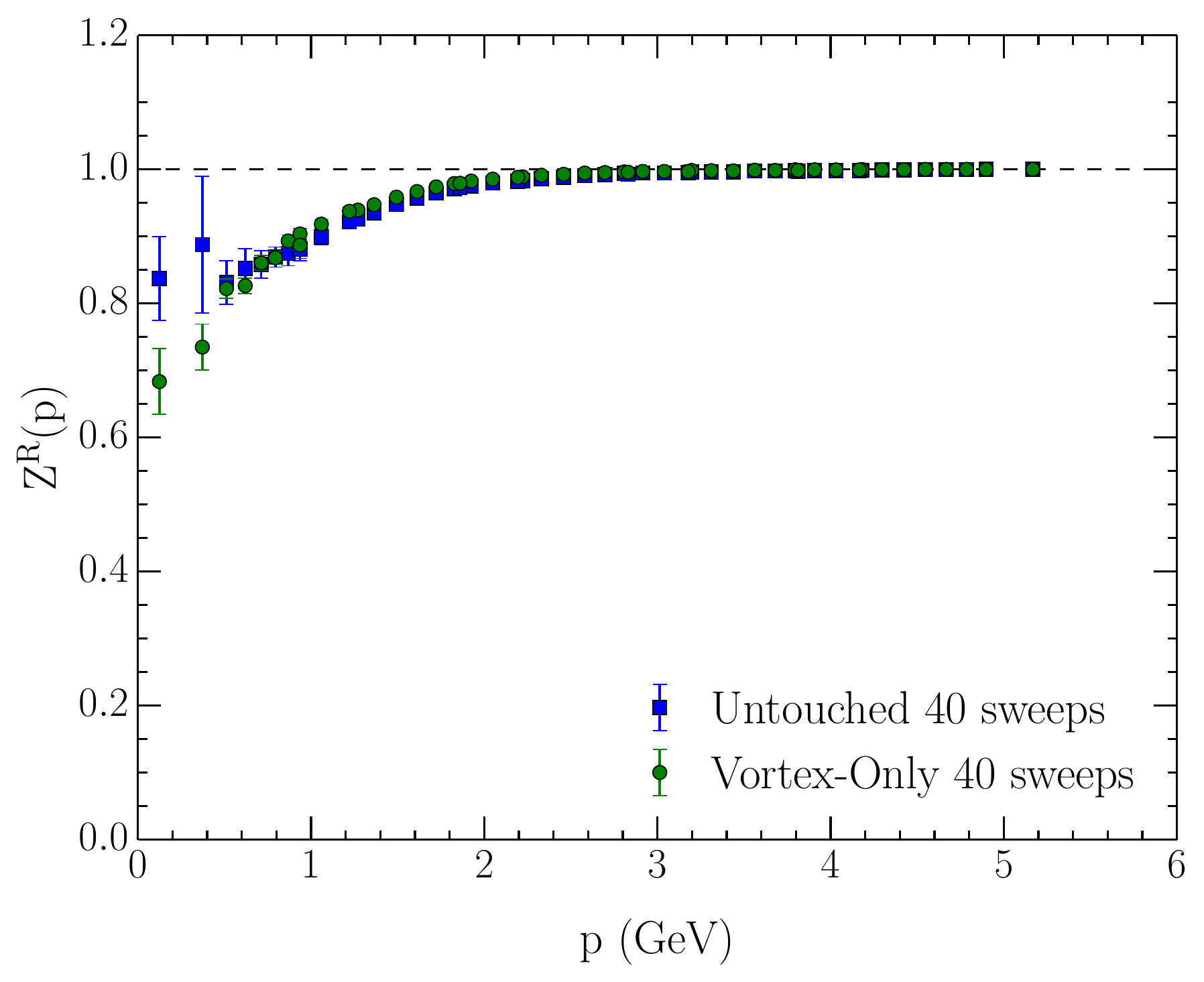}}
\subfigure[]{
\label{M02200UTc40VOc40}
\includegraphics[width=\columnwidth]{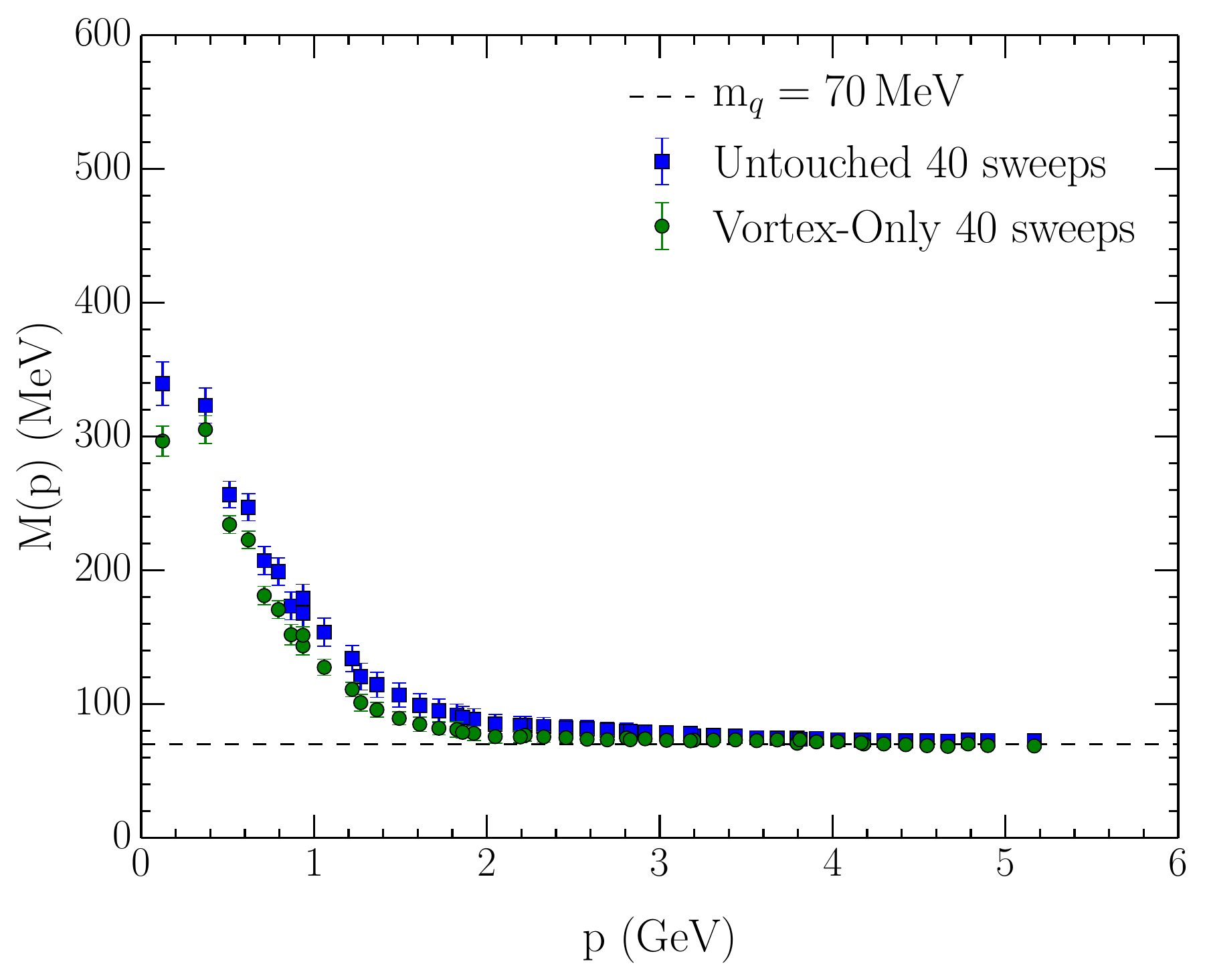}}
\subfigure[]{
\label{Z02200UTc40VOc40}
\includegraphics[width=\columnwidth]{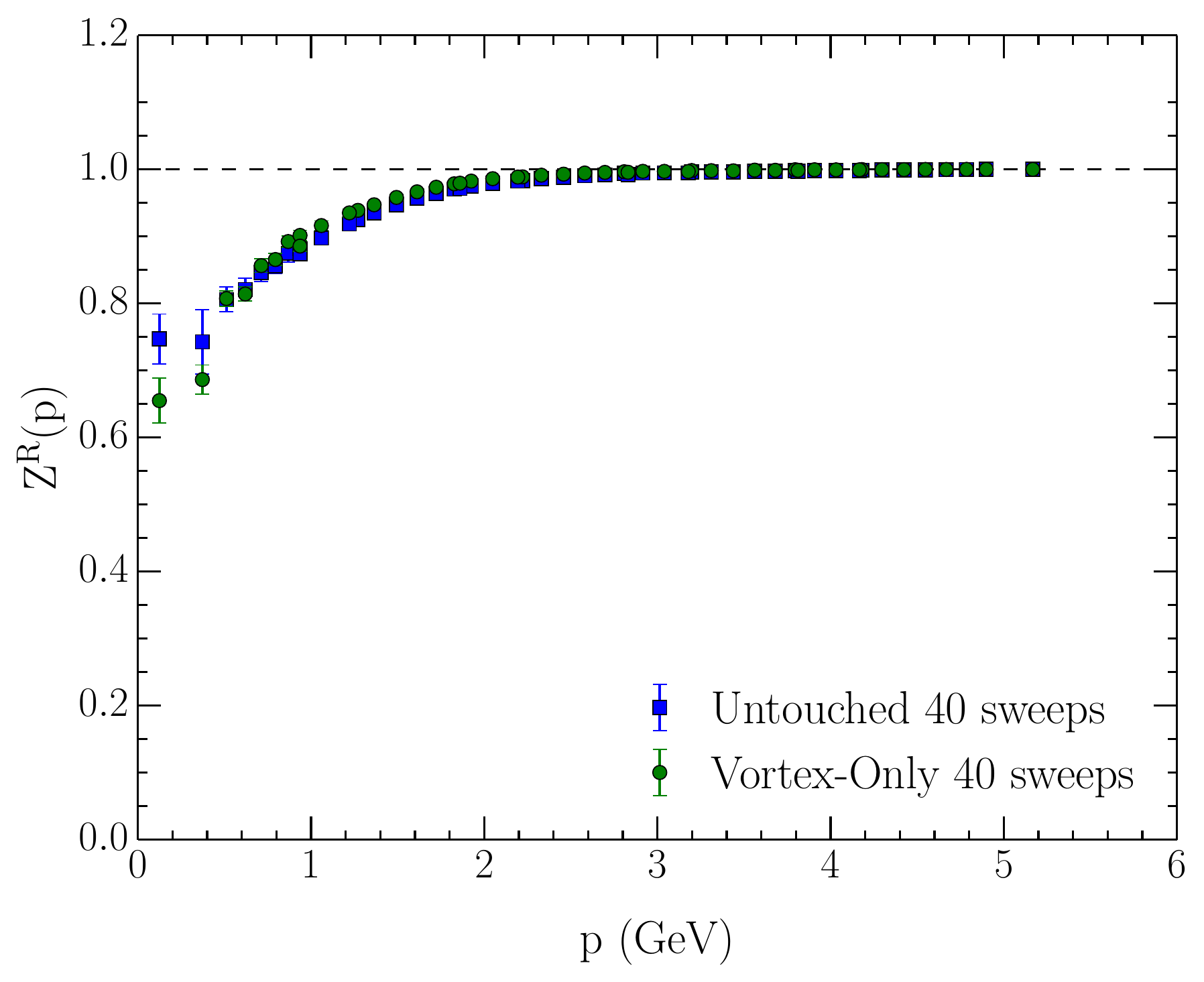}}
\caption{The mass (left) and renormalization (right) functions on untouched (blue squares), and vortex-only (green circles) ensembles after 40 sweeps of cooling, at physical bare quark masses of $12$ MeV (\subref{M00400UTc40VOc40} \& \subref{Z00400UTc40VOc40}), $40$ MeV (\subref{M01200UTc40VOc40} \& \subref{Z01200UTc40VOc40}), and $70$ MeV (\subref{M02200UTc40VOc40} \& \subref{Z02200UTc40VOc40}).} 
\label{Fig:MUTc40VOc40}
\end{figure*}
\begin{figure*}[thpb]
\subfigure[]{
\label{M00400UTc80VOc80}
\includegraphics[width=\columnwidth]{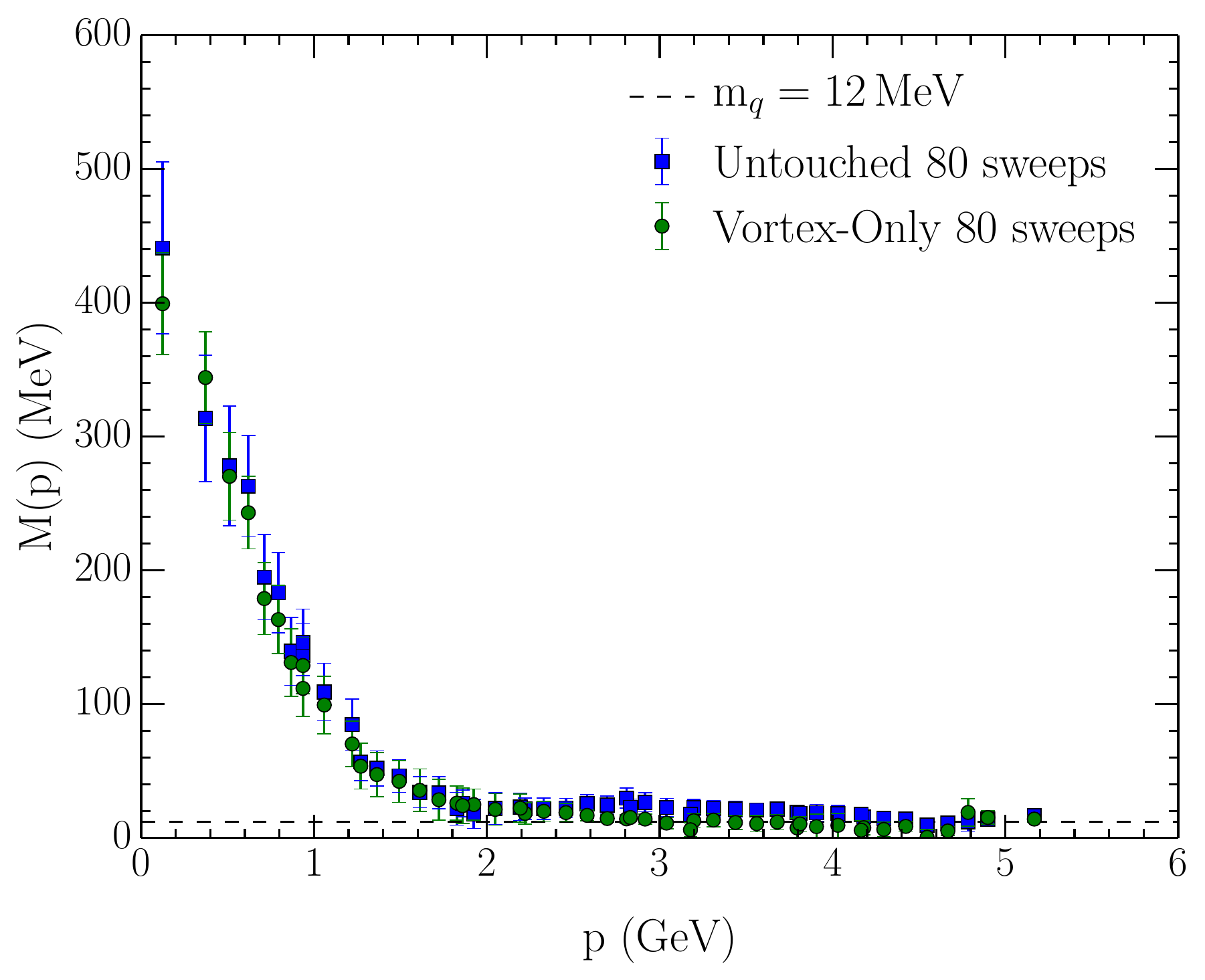}}
\subfigure[]{
\label{Z00400UTc80VOc80}
\includegraphics[width=\columnwidth]{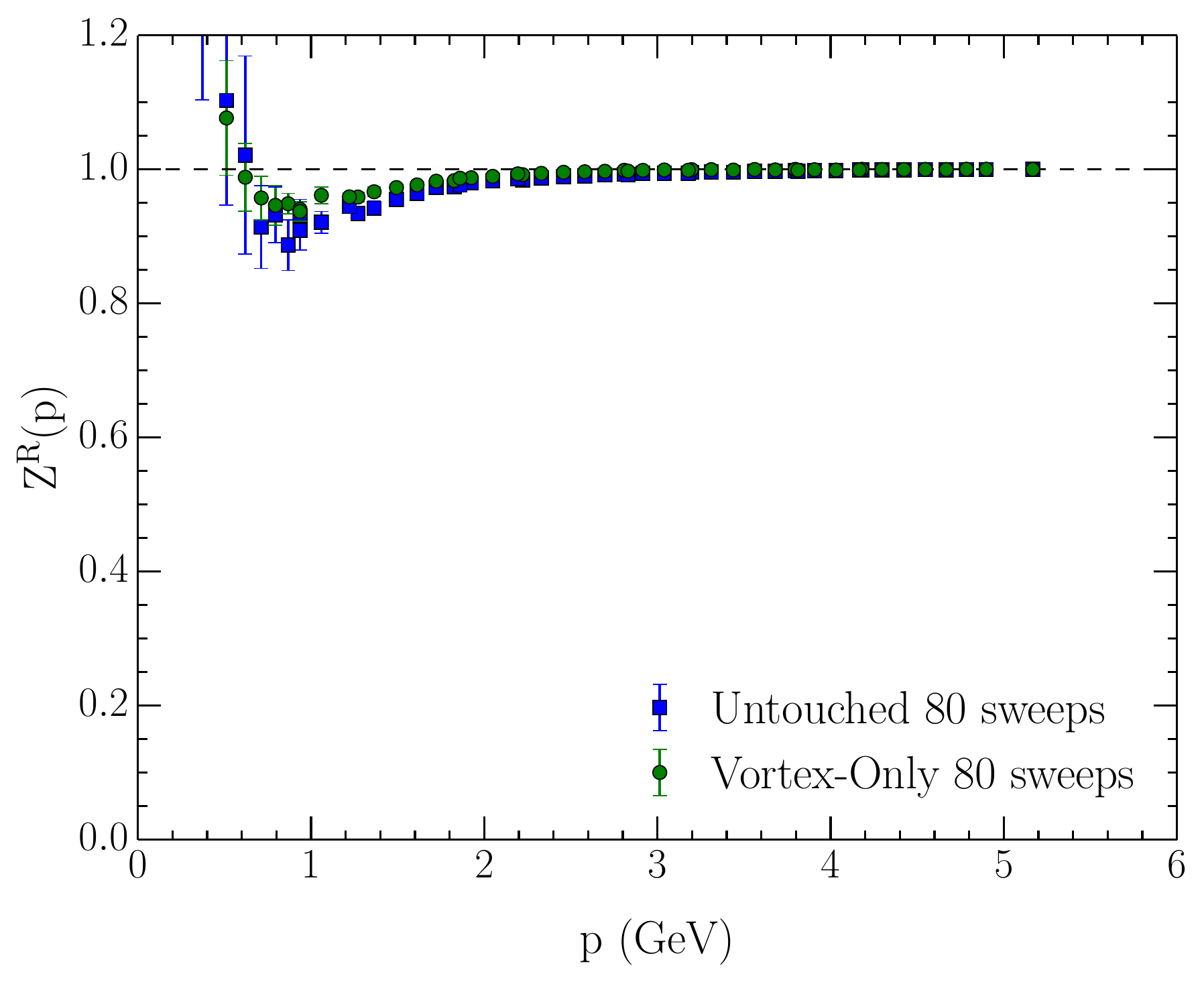}}
\subfigure[]{
\label{M01200UTc80VOc80}
\includegraphics[width=\columnwidth]{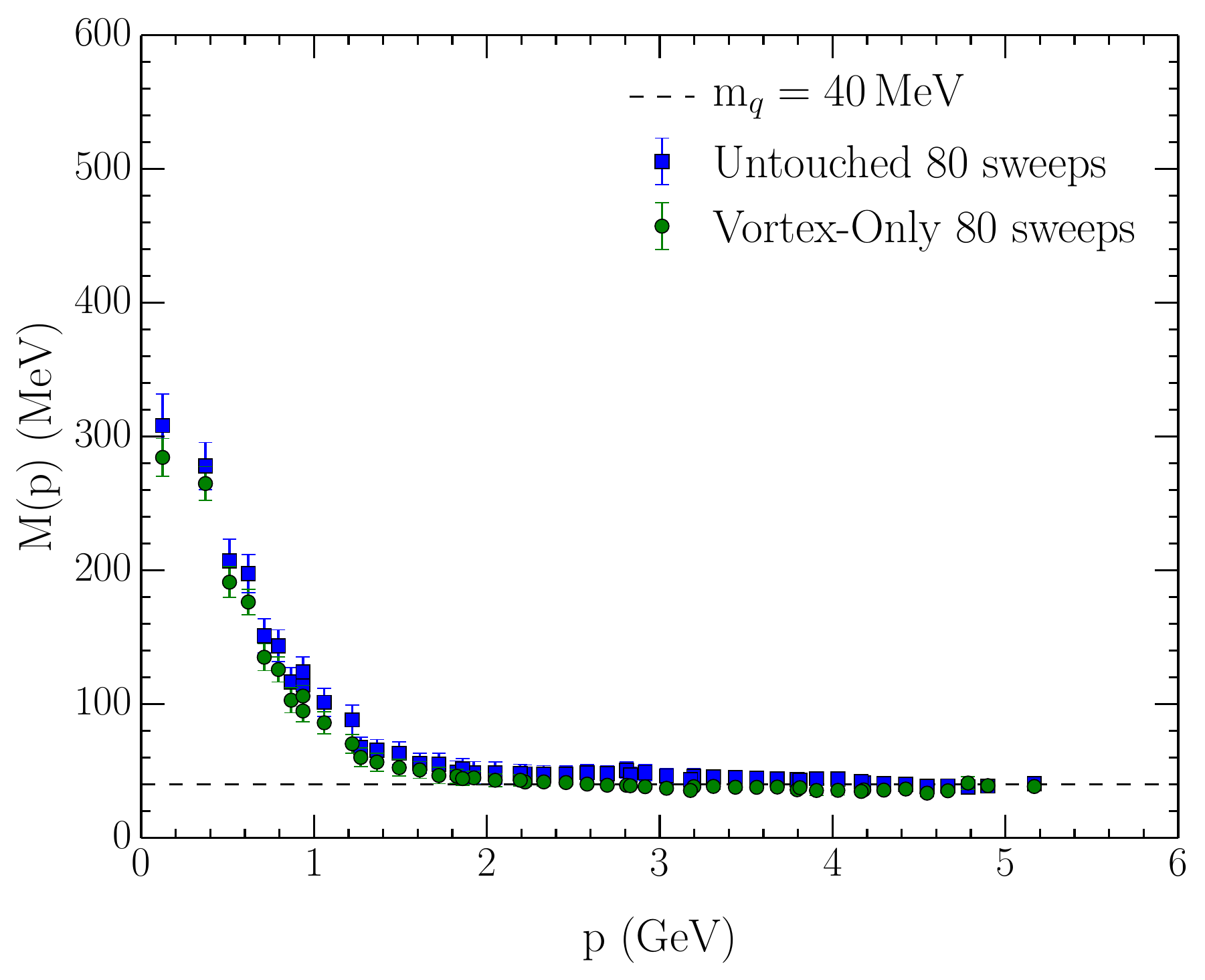}}
\subfigure[]{
\label{Z01200UTc80VOc80}
\includegraphics[width=\columnwidth]{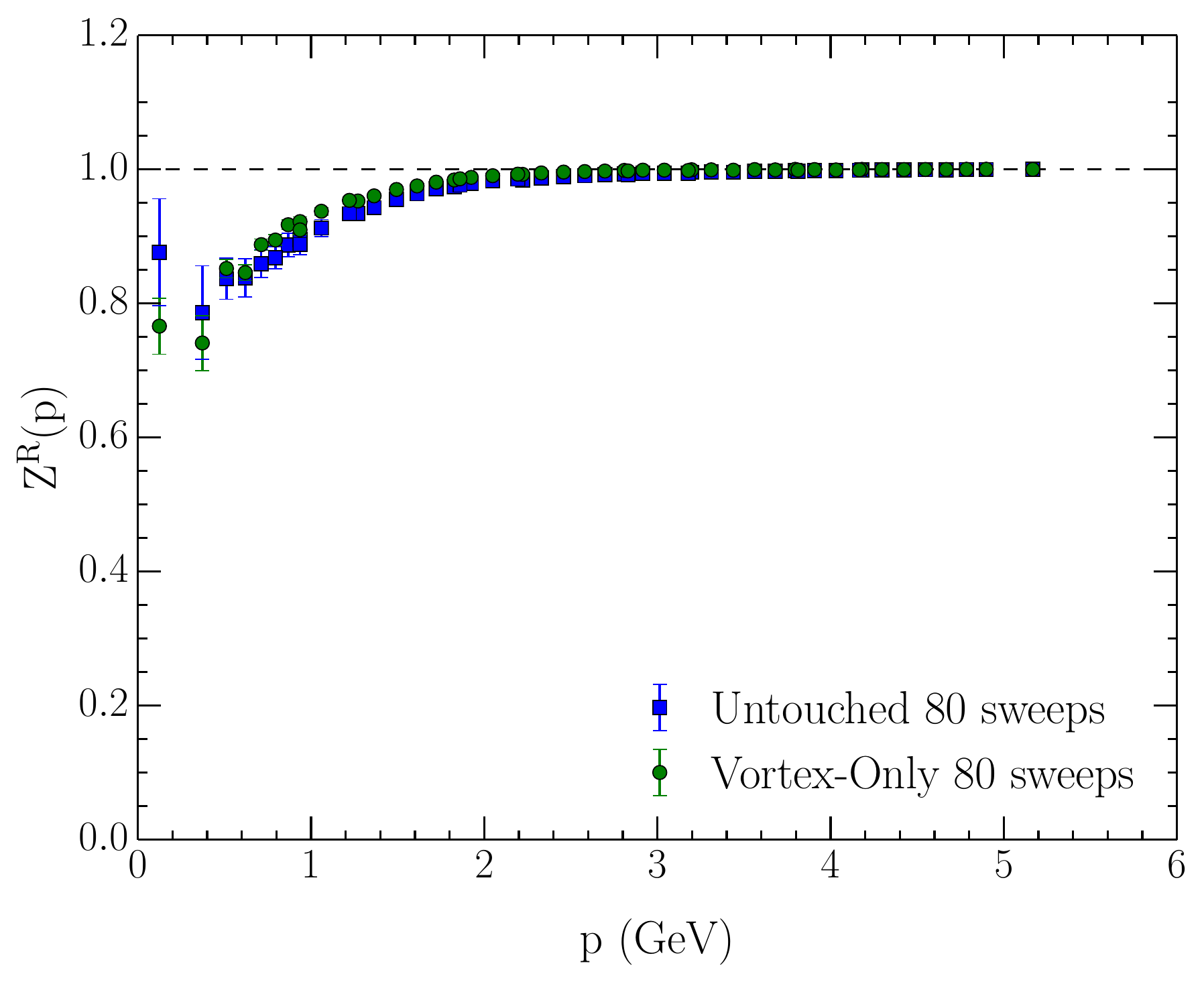}}
\subfigure[]{
\label{M02200UTc80VOc80}
\includegraphics[width=\columnwidth]{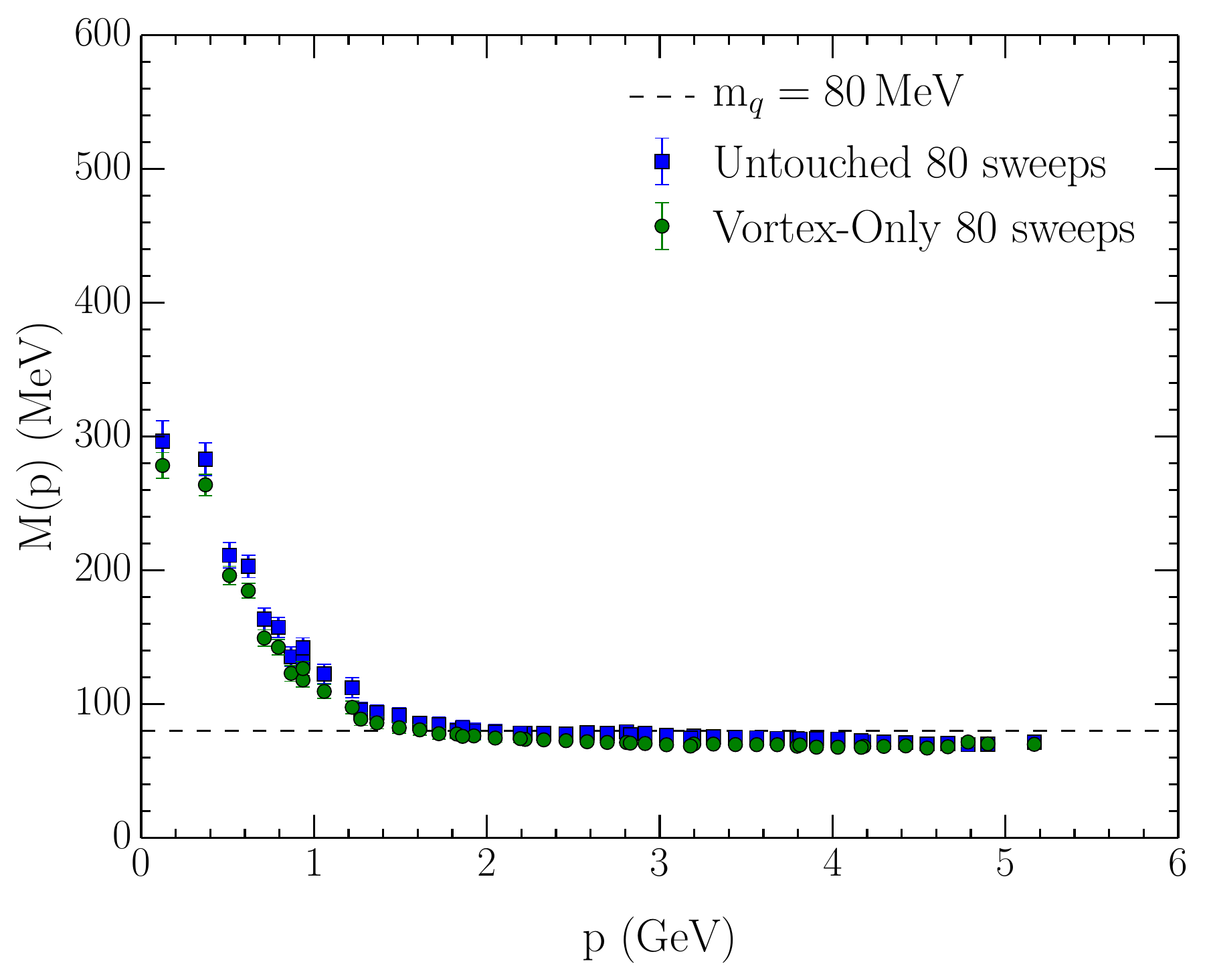}}
\subfigure[]{
\label{Z02200UTc80VOc80}
\includegraphics[width=\columnwidth]{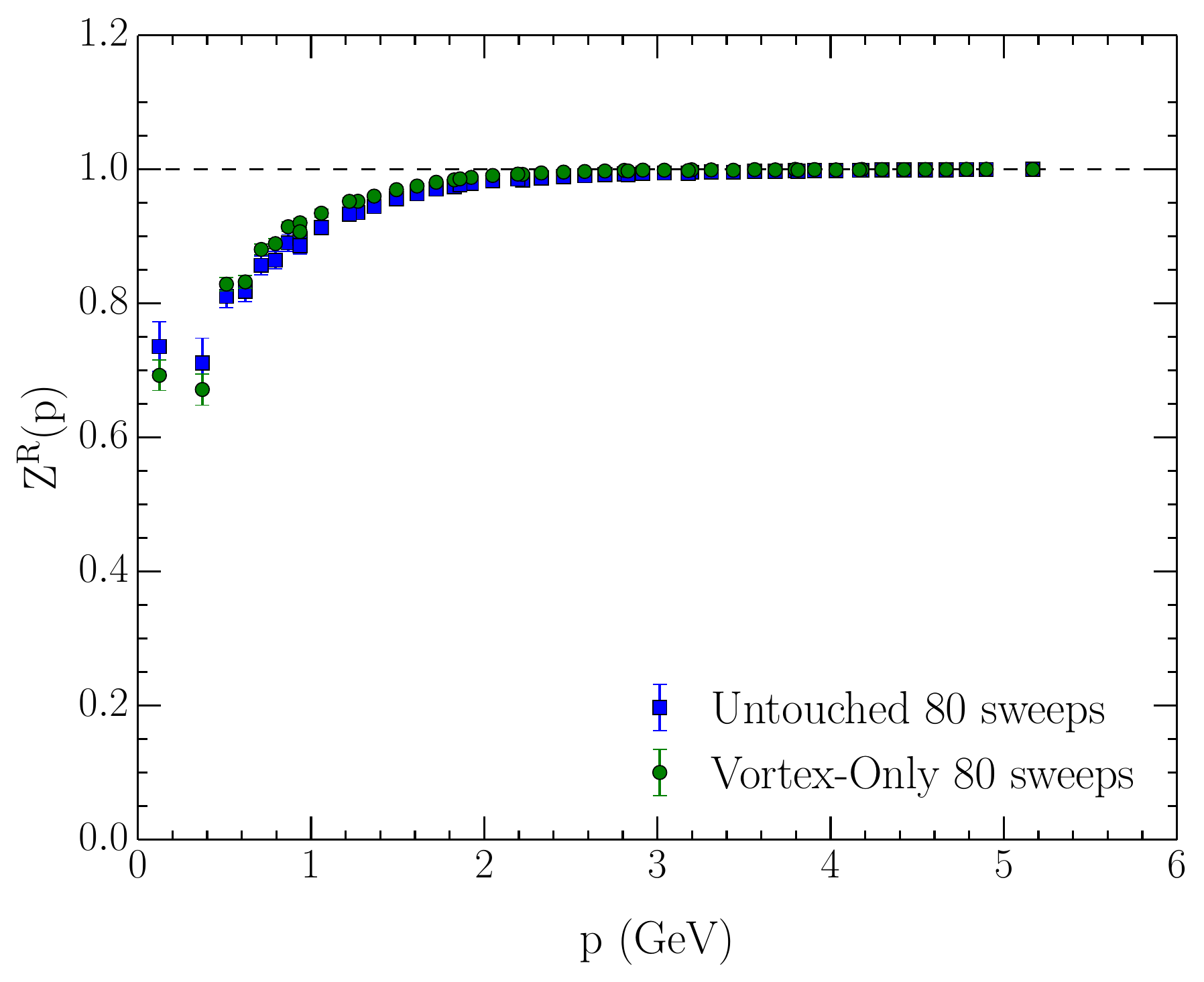}}
\caption{The mass and renormalization functions on untouched (blue squares), vortex-only (green circles) ensembles after 80 sweeps of cooling, at physical bare quark masses of $12$ MeV (\subref{M00400UTc80VOc80} \& \subref{Z00400UTc80VOc80}), $40$ Mev (\subref{M01200UTc80VOc80} \& \subref{Z01200UTc80VOc80}), and $70$ MeV (\subref{M02200UTc80VOc80} \& \subref{Z02200UTc80VOc80}).} 
\label{Fig:MUTc80VOc80}
\end{figure*}
Results are presented in Figs.~\ref{Fig:MUTc40VOc40}, and \ref{Fig:MUTc80VOc80}. As expected, the mass function in the untouched case displays strong infrared enhancement; dynamical mass generation via dynamical chiral symmetry breaking. The cooling procedure, which has left the instanton structure of the configurations intact, has correspondingly retained dynamical mass generation on cooled configurations. Although there is a small decrease in the amount of dynamically generated mass from 40 to 80 sweeps, due to pair annihilation of instantons, the qualitative long-range features of the quark propagator remain intact. At short range, cooling has removed all UV noise, and so consequently the mass function sits on the input bare mass at large momenta.\par
The renormalisation function on the untouched ensemble likewise displays the characteristic shape; dipping slightly in the IR, reaching a plateau in the UV. These observations remain true at all bare quark masses considered. \par
As found in Ref.~\cite{Trewartha:2015nna}, the vortex-only results are able to reproduce both the mass and renormalisation functions of the untouched ensembles. The background of instanton-like objects found after smoothing on vortex-only configurations has successfully reproduced the features of the quark propagator. This concords with the results of Ref.~\cite{Ilgenfritz:2008ia}, which showed the overlap operator displaying a similar gauge field background to that found after smoothing. Although it requires a higher level of smoothing for topological objects to become apparent in the gauge field structure itself, the overlap is sensitive to their presence after just 10 sweeps. We have shown the equivalence of vortex-only and untouched results persisting up to a high level of smoothing, and at high bare quark masses. The similar background of instanton-like objects on vortex-only and untouched ensembles observed by directly examining the gauge fields equally produces dynamical mass generation on both untouched and vortex-only configurations.

\section{Conclusions}
\label{sec:conc}

We have examined the gauge field structure of configurations consisting solely of centre vortices under smoothing, comparing to the original configurations under the same smoothing. Through study of the action, integrated topological charge, and instanton content of these configurations, we have found remarkably similar structures emerging under smoothing, regardless of whether cooling or over-improved stout-link smearing is used. In both cases, after smoothing we obtain a gauge field structure approximating a background of instantons, stable under further smoothing. Under cooling, instanton-like objects begin to appear on the vortex-only configurations after 10 sweeps, and after 40 sweeps the gauge field background of both the vortex-only and untouched ensembles consist of a dilute instanton liquid.

This provides a mechanism for the findings of Ref.~\cite{Trewartha:2015nna}, that after 10 sweeps of cooling vortex-only configurations support dynamical mass generation, and thus dynamical chiral symmetry breaking. The information contained in the centre vortex degree of freedom is sufficient to recreate the underlying long-range structure of the QCD vacuum, revealed through smoothing. We have also confirmed that the gauge field background of vortex-removed configurations is unstable under smoothing. \par 
Upon examining the string tension, we have found that vortex-only configurations are, again, capable of reproducing the untouched results under smoothing. The similar long-range gauge field structure on vortex-only and untouched configurations after cooling has produced a similar confining structure, as well as similar dynamical mass generation.  \par
Although we have found similar gauge field background structure on the untouched and vortex-only configurations after smearing, this naturally has removed the short-range information of the gauge fields. It would be desirable to have a method of smoothing that would allow us to gain smooth centre-vortex configurations without removing relevant short-range structure, allowing direct comparison to unsmeared configurations. It would be interesting to explore an $\SU{3}$ extension to the technique proposed for the $\SU{2}$ case in Ref.~\cite{Hollwieser:2015koa}.

We have demonstrated how gauge-field smoothing can restore agreement
between the characteristic features of confinement and dynamical chiral
symmetry breaking observed on untouched and vortex-only ensembles. There
are two possible interpretations of this result. In the first scenario
we can consider that the discrepancies are associated with gauge-field
roughness which is removed under smearing or cooling, thereby restoring
agreement. An alternative interpretation is  that the essential
fundamental feature of the QCD vacuum is the thick centre vortex, rather
than the thin centre vortices that are identified on the lattice. Then,
smoothing would be primarily understood as a process of creating thick
centre vortices from the identified thin centre vortices. The
characteristic size of thick vortices is around $1\,\,\mathrm{fm}$; we
have found that 10 sweeps of cooling, corresponding to a cooling radius
of $1.2\,\,\mathrm{fm}$, is sufficient to recreate dynamical chiral
symmetry breaking from a centre vortex background. It is of note that
while vortex-only configurations consist only of thin centre vortices,
the vortex removal procedure is sufficient to remove thick centre
vortices from the lattice. This would explain the result that while
vortex removal is sufficient to spoil dynamical chiral symmetry breaking
and confinement, smearing is required to reproduce these from
vortex-only configurations.

The two interpretations we have postulated can be distinguished by their
behaviour under the approach to the continuum limit. Under the first
scenario, to ameliorate the roughness of the vortex-only fields it would
be sufficient to hold the number of smoothing sweeps fixed as the
lattice spacing tends to zero. Under the second scenario, it would be
necessary to increase the number of smoothing sweeps used as the lattice
spacing decreases, such that the physical smearing distance is fixed to
preserve the scale corresponding to the characteristic thickness of
centre vortices. This will be the subject of future work.

\section*{Acknowledgements}
This research was undertaken with the assistance of resources awarded at the NCI National Facility in Canberra, Australia. This work was supported by resources provided by the Pawsey Supercomputing Centre with funding from the Australian Government and the Government of Western Australia. These resources are provided through the National Computational Merit Allocation Scheme and the University of Adelaide Partner Share supported by the Australian Government.
This research is supported by the Australian Research Council through
grants DP120104627, DP150103164, LE120100181 and LE110100234.
\vspace{0.5cm}
\appendix*
\section{MCG fixing coefficients}
\label{Sec:appendix}

For the Maximal Centre Gauge fixing procedure, we wish to maximize the quantity
\begin{eqnarray}
R_{\textrm{local}}(x') = &\sum_{\mu}&|\, \trace \, \Omega(x')\,U_{\mu}(x')|^{2} \,+\,  \\ \nonumber
&\sum_{\mu}&|\, \trace \, U_{\mu}(x' - \hat{\mu})\,\Omega^{\dagger}(x')|^{2},
\end{eqnarray}
at a given lattice site $x'$, for a local gauge transformation $\Omega(x')$.

This is achieved by restricting $\Omega(x')$ to be in one of the $\SU{2}$ subgroups of $\SU{3}$. Then, we parametrize an $\SU{2}$ matrix $[\Omega(x')]_{\SU{2}}$ by
\begin{equation}
[\Omega(x')]_{\SU{2}} = \Omega_{4}\,\identity - i\Omega_{i}\,\sigma_{i},
\end{equation}
and embed the resulting $\SU{2}$ matrix in one of the 3 $\SU{2}$ subgroups of $\SU{3}$. This allows us to rewrite the gauge fixing condition in terms of the $\SU{2}$ parametrization as
\begin{equation}
R_{\textrm{local}}(x') = \sum_{i,j=1}^{4} \frac{1}{2} \Omega_{i}\,a_{ij}\,\Omega_{j} - \sum_{i}^{4}\Omega_{i}\,b_{i} + c.
\end{equation}
Where $a_{ij}$ are elements of a real, symmetric matrix, $b_{i}$ a real vector and $c$ a real constant, all dependent only on $U_{\mu}(x')$ and $U_{\mu}(x' - \hat{\mu})$. Note that we suppress the sum over directions, $\mu$. Once the coefficients are defined the local quantity $R_{\textrm{local}}(x')$ can be maximised according to the method of, e.g., Ref.~\cite{QCQPpaper}.

The gauge fixing coefficients $a_{ij}$, $b_{i}$, and $c$ are defined in terms of the gauge links $U_{\mu}(x')$ and $U_{\mu}(x' - \hat{\mu})$, for any given $\SU{2}$ subgroup of $\SU{3}$.

It is convenient to define
\begin{eqnarray}
U \, &=& \,U_{\mu}(x')\,, \nonumber \\
V \, &=& \,U_{\mu}(x - \hat{\mu})\,, \nonumber \\
\end{eqnarray}
Then we define complex $2 \times 2$ matrices, $\UM$ and $\VM$, as the components of $U$ and $V$ corresponding to the given $\SU{2}$ subgroup, and complex scalars, $\US$ and $\VS$, as the remaining diagonal element. The other elements of $U$ and $V$ do not contribute to the coefficients, and so for convenience are set to $0$.

As an example, for the first $\SU{2}$ subgroup, we set
\begin{equation}
U = \left( \begin{array}{cc}
\UM & \begin{array}{cc} 0 \\ 0 \end{array} \\
\begin{array}{cc} 0 & 0 \end{array} & \US \\
\end{array} \right)\,,
\end{equation}
and
\begin{equation}
V = \left( \begin{array}{cc}
\VM & \begin{array}{cc} 0 \\ 0 \end{array} \\
\begin{array}{cc} 0 & 0 \end{array} & \VS \\
\end{array} \right)\,.
\end{equation}

Once we have defined $\UM$, $\US$, $\VM$, and $\VS$, the values of the gauge-fixing coefficients depend only on these, and so the expressions below are valid for all $\SU{2}$ subgroups.

\begin{widetext}
Then we have for the real symmetric matrix $a_{ij}$,
\begin{eqnarray}
a_{11} &=& 
 \,|\UM_{12}|^{2}\,
+\,|\UM_{21}|^{2}\,
+\,\UM_{12}\,\UMC_{21}\, 
+\,\UM_{21}\,\UMC_{12}\,
+\,|\VM_{12}|^{2}\,
+\,|\VM_{21}|^{2}\,
+\,\VM_{12}\VMC_{21}\, 
+\,\VM_{21}\VMC_{12}\, 
, \\
a_{12} &=& 2\, i\, \left (
 \,\UM_{12}\UMC_{21} \,
-\,\UM_{21}\UMC_{12} \,
+\,\VM_{12}\VMC_{21} \,
-\,\VM_{21}\VMC_{12} \, 
\right ) \, ,
\\
a_{13} &=& 
+\,\UM_{11}\UMC_{12}\,
+\,\UM_{12}\UMC_{11}\,
+\,\UM_{11}\UMC_{21}\,
+\,\UM_{21}\UMC_{11}\,
-\,\UM_{22}\UMC_{12}\,
-\,\UM_{12}\UMC_{22}\,
-\,\UM_{22}\UMC_{21}\,
-\,\UM_{21}\UMC_{22}\,
\nonumber \\
&& 
+\,\VM_{11}\VMC_{12}\,
+\,\VM_{12}\VMC_{11}\,
+\,\VM_{11}\VMC_{21}\,
+\,\VM_{21}\VMC_{11}\,
-\,\VM_{22}\VMC_{12}\,
-\,\VM_{12}\VMC_{22}\,
-\,\VM_{22}\VMC_{21}\,
-\,\VM_{21}\VMC_{22}\, 
, \\
a_{14} &=& i\, \bigl (
 \,\UM_{11}\UMC_{12}\,
-\,\UM_{12}\UMC_{11}\,
+\,\UM_{11}\UMC_{21}\,
-\,\UM_{21}\UMC_{11}\,
+\,\UM_{22}\UMC_{12}\,
-\,\UM_{12}\UMC_{22}\,
+\,\UM_{22}\UMC_{21}\,
-\,\UM_{21}\UMC_{22}\,
 \nonumber \\
&&\; 
-\,\VM_{11}\VMC_{12}\,
+\,\VM_{12}\VMC_{11}\,
-\,\VM_{11}\VMC_{21}\,
+\,\VM_{21}\VMC_{11}\,
-\,\VM_{22}\VMC_{12}\,
+\,\VM_{12}\VMC_{22}\,
-\,\VM_{22}\VMC_{21}\,
+\,\VM_{21}\VMC_{22}\,
\bigr ) \, ,
 \\
a_{22} &=& 
 \,|\UM_{12}|^{2}\,
+\,|\UM_{21}|^{2}\,
-\,\UM_{12}\UMC_{21}\,
-\,\UM_{21}\UMC_{12}\,
+\,|\VM_{12}|^{2}\,
+\,|\VM_{21}|^{2}\,
-\,\VM_{12}\VMC_{21}\,
-\,\VM_{21}\VMC_{12}\, 
, \\
a_{23} &=& i\, \bigl (
-\,\UM_{11}\UMC_{12}\,
+\,\UM_{12}\UMC_{11}\,
+\,\UM_{11}\UMC_{21}\,
-\,\UM_{21}\UMC_{11}\, 
+\,\UM_{22}\UMC_{12}\,
-\,\UM_{12}\UMC_{22}\,
-\,\UM_{22}\UMC_{21}\,
+\,\UM_{21}\UMC_{22}\,
\nonumber \\
&&\quad
-\,\VM_{11}\VMC_{12}\,
+\,\VM_{12}\VMC_{11}\,
+\,\VM_{11}\VMC_{21}\, 
-\,\VM_{21}\VMC_{11}\,
+\,\VM_{22}\VMC_{12}\,
-\,\VM_{12}\VMC_{22}\,
-\,\VM_{22}\VMC_{21}\,
+\,\VM_{21}\VMC_{22}\,
\bigr ) \, ,
\\
a_{24} &=& 
+\,\UM_{11}\UMC_{12}\,
+\,\UM_{12}\UMC_{11}\,
-\,\UM_{11}\UMC_{21}\,
-\,\UM_{21}\UMC_{11}\,
+\,\UM_{22}\UMC_{12}\,
+\,\UM_{12}\UMC_{22}\,
-\,\UM_{22}\UMC_{21}\,
-\,\UM_{21}\UMC_{22}\,
 \nonumber \\
&& 
-\,\VM_{11}\VMC_{12}\,
-\,\VM_{12}\VMC_{11}\,
+\,\VM_{11}\VMC_{21}\,
+\,\VM_{21}\VMC_{11}\,
-\,\VM_{22}\VMC_{12}\,
-\,\VM_{12}\VMC_{22}\,
+\,\VM_{22}\VMC_{21}\,
+\,\VM_{21}\VMC_{22}\,
, \\
a_{33} &=& 
 \,|\UM_{11}|^{2}\,
+\,|\UM_{22}|^{2}\,
-\,\UM_{11}\UMC_{22}\,
-\,\UM_{22}\UMC_{11}\,
+\,|\VM_{11}|^{2}\,
+\,|\VM_{22}|^{2}\,
-\,\VM_{11}\VMC_{22}\,
-\,\VM_{22}\VMC_{11} 
, \\
a_{34} &=& 2 \, i\, \left (
-\,\UM_{11}\UMC_{22}\,
+\,\UM_{22}\UMC_{11}\,  
+\,\VM_{11}\VMC_{22}\,
-\,\VM_{22}\VMC_{11}\,
\right ) \, ,
\\
a_{44} &=& 
 \,|\UM_{11}|^{2}\,
+\,|\UM_{22}|^{2}\,
+\,\UM_{11}\UMC_{22}\,
+\,\UM_{22}\UMC_{11}\,
+\,|\VM_{11}|^{2}\,
+\,|\VM_{22}|^{2}\,
+\,\VM_{11}\VMC_{22}\,
+\,\VM_{22}\VMC_{11}\, 
,
\end{eqnarray}
and for the real vector $b_{i}$,
\begin{eqnarray}
b_{1} &=& i\, \left (
-\, \UM_{12}\,\USC\,
+\,\UMC_{12}\,\US\,
-\, \UM_{21}\,\USC\,
+\,\UMC_{21}\,\US\,
+\, \VM_{12}\,\VSC\,
-\,\VMC_{12}\,\VS\,
+\, \VM_{21}\,\VSC\,
-\,\VMC_{21}\,\VS\, 
\right ) \, ,
\\
b_{2} &=&\qquad \!
 \, \UM_{12}\,\USC\,
+\,\UMC_{12}\,\US\,
-\, \UM_{21}\,\USC\,
-\,\UMC_{21}\,\US\,
-\, \VM_{12}\,\VSC\,   
-\,\VMC_{12}\,\VS\,
+\, \VM_{21}\,\VSC\,
+\,\VMC_{21}\,\VS\, 
, \\
b_{3} &=& i\, \left (
-\, \UM_{11}\,\USC\,
+\,\UMC_{11}\,\US\,
+\, \UM_{22}\,\USC\,
-\,\UMC_{22}\,\US\,
+\, \VM_{11}\,\VSC\,
-\,\VMC_{11}\,\VS\,
-\, \VM_{22}\,\VSC\,
+\,\VMC_{22}\,\VS\, 
\right ) \, ,
\\
b_{4} &=& \qquad \!
 \, \UM_{11}\,\USC\,
+\,\UMC_{11}\,\US\,
+\, \UM_{22}\,\USC\,
+\,\UMC_{22}\,\US\,
+\, \VM_{11}\,\VSC\,
+\,\VMC_{11}\,\VS\,
+\, \VM_{22}\,\VSC\,
+\,\VMC_{22}\,\VS \, 
,
\end{eqnarray}
and
\begin{equation}
c = |\US|^{2}\,+\,|\VS|^{2} 
\, .
\end{equation}

\end{widetext}
\bibliography{PRD270115}
\end{document}